\documentclass[a4paper,pra,superscriptaddress,floatfix,nofootinbib,twocolumn]{revtex4-1}

\usepackage{bbm}
\usepackage[pdftex]{graphicx}
\usepackage{latexsym,amsmath,verbatim,amssymb,txfonts, mathtools}
\usepackage{color}
\usepackage{rotating}
\usepackage{verbatim}
\usepackage{multirow}
\usepackage[english]{babel}
\usepackage{comment}
\usepackage{hyperref}
\usepackage[normalem]{ulem}
\usepackage{tabularx}
\usepackage{braket}

\newcommand{\EM}{Appendix}

\newcommand{\change}[1]{#1}
\newcommand{\newchange}[1]{#1}

\begin{document}

\title{Enabling quantum communication in ultra-large-scale networks}

\author{Filippo Radicchi}
\affiliation{Center for Complex Networks and Systems Research, Luddy School
  of Informatics, Computing, and Engineering, Indiana University, Bloomington,
  Indiana 47408, USA}
  \email{f.radicchi@gmail.com}

\begin{abstract}
The recent development of small-scale quantum networks poses the question of
\newchange{which network-level mechanisms could enable quantum communication to
operate at scale in a future Quantum Internet}.
The question can be answered within a well-established framework where a quantum
network is mapped into a classical weighted graph, and communication reliability
is assessed using methods from network theory.
Previous work relied on this framework to determine sufficient conditions for
viable network-wide communication only on special network topologies, such as
regular lattices, and to devise communication protocols applicable only to
small- and moderate-sized real networks.
Here, we overcome both limitations by introducing a family of quantum
communication protocols that scale to networks with hundreds of millions of
nodes\newchange{. We show that the competition between increasing communication
distance and path redundancy determines their large-scale behavior, and that
degree heterogeneity can provide sufficient redundancy to sustain communication
in small-world networks}.
These conclusions are supported by a systematic analysis of both
\newchange{real network topologies} and synthetic networks, and
\newchange{an analytical treatment} in the limit of infinitely large random
scale-free graphs.
\end{abstract}

\maketitle

\section{Introduction}
Sustaining network-wide communication at scale is the key to the success of the Internet, a human-made infrastructure that grew from the tiny ARPANET with just four nodes in 1969~\cite{arpanet} to a massive network composed of almost hundreds of thousands of autonomous systems and more than twenty billion devices in 2025~\cite{as, iot}. Such a capability is enabled not only by advances in hardware and software, but is also supported by a distinctive network structure. Being the result of the collective effort of many individuals without a centralized control, the Internet is in fact a scale-free, small-world network similar to many other self-organized networks observed in nature~\cite{faloutsos1999power, watts1998collective, barabasi1999emergence}. From this perspective, microscopic details, such as the specifics of how computers work and communicate, are not essential to understand macroscopic, system-wide properties of the network~\cite{pastor2004evolution}, as for example its growth~\cite{barabasi1999emergence, pastor2001dynamical, vazquez2002large, papadopoulos2012popularity}, its resilience to damage~\cite{cohen2000resilience, albert2000error, callaway2000network}, its vulnerability to the diffusion of viruses~\cite{pastor2001epidemic, moreno2002epidemic, newman2002spread, chakrabarti2008epidemic}, as well as its adaptability and ability to support searching and routing~\cite{kleinberg2000navigation, watts2002identity, adamic2001search, boguna2009navigability}.

%Present of quantum communication networks
A breakthrough comparable to the Internet will likely occur with the advent of the Quantum Internet, i.e., the infrastructure that will leverage the principles of quantum mechanics to enable a new frontier of communication~\cite{kimble2008quantum, wehner2018quantum}. Although the theoretical foundations of quantum communication are older than ARPANET~\cite{gordon1962quantum}, the actual realization of quantum communication networks is literally occurring at the time of writing~\cite{quant-net, bersin2024development, NYSQIT, IEQNET, NG-QNet, rahmouni2024100, oakridge, q-next, EuroQCI, chen2021integrated}.
These small-scale networks represent the precursors to  an infrastructure with unprecedented capabilities. Distributed quantum sensing~\cite{zhang2021distributed}, distributed quantum computing~\cite{boschero2025distributed}, and information-theoretically secure communication~\cite{zhang2025towards} are among the anticipated applications of the Quantum Internet.  However, future uses of the infrastructure cannot be fully predicted, just as the impact of the Internet could not have been foreseen when the first classical networks were created. 

The range of potential applications of the Quantum Internet  will crucially depend on its ability to sustain large-scale network-wide communication.
In a quantum network, information propagates from one node to another thanks to the distribution of entanglement~\cite{rohde2021quantum}. 
\newchange{The physical realization of long-distance entanglement distribution
requires overcoming effects such as transmission loss and noise, motivating
the development of quantum-repeater architectures~\cite{azuma2023quantum}.}
\change{Physically realistic simulations of entanglement distribution are computationally demanding, limiting their application to the analysis of small- and moderate-sized networks~\cite{Chakraborty2020,Vardoyan2024,Caleffi2017,Satoh2022}.}
\newchange{Network-level approaches have addressed complementary aspects of
the problem, including the statistical properties of quantum networks
generated from physically motivated models~\cite{brito2020statistical} and
the minimization of resource costs for distributing entanglement through
quantum networks~\cite{azuma2025networking}.}
By introducing an additional level of abstraction, entanglement distribution can be modeled within a classical network framework~\cite{acin2007entanglement}.
%%%
\newchange{This framework provides a coarse-grained description of quantum
communication aimed at isolating the role of network topology. It does not
explicitly account for microscopic effects such as dynamical noise,
decoherence, finite memory times, and implementation-dependent imperfections.
At the same time, it does not assume perfect elementary entanglement, as the
quality of the entangled states associated with individual edges can vary
continuously. The abstraction therefore makes it possible to investigate how
network topology interacts with elementary entanglement quality at scales that
are inaccessible to more detailed physical descriptions.}
%%%
%The framework does not explicitly model some physical aspects of quantum communication systems, e.g., noise and decoherence. However, it enables the analysis of much larger systems and complements traditional approaches by highlighting the role of network topology in quantum communication.}
%
Still, current results that take advantage of such a classical network framework are limited to
special topologies like regular and Bethe lattices, and random annealed networks~\cite{acin2007entanglement, cuquet2009entanglement,  perseguers2010quantum, perseguers2010multipartite, perseguers2013distribution, siomau2016quantum, khanna2024quantum, meng2021concurrence, malik2022concurrence, meng2023percolation, hu2025unveiling, wang2026counter}.
\newchange{Within this framework, communication protocols applicable to
real network topologies have so far been investigated only for
relatively small networks}~\cite{degirolamo2025percolation}.
%Also, no practical communication protocols have been developed so far for real network topologies, if not for relatively small networks~\cite{degirolamo2025percolation}. 
%It is therefore unclear whether the Quantum Internet will be ever able to sustain a growth comparable to the one of its classical counterpart.
\newchange{It is therefore unclear whether quantum communication remains viable
on general network topologies at increasingly large scales.}
 
%In this paper, we offer a positive resolution to this doubt by showing that quantum communication is viable on ultra-large-scale networks, \change{as long as the networks have a degree-heterogeneous and small-world structure like the Internet~\cite{faloutsos1999power, watts1998collective, barabasi1999emergence}.} 
%\change{We reach this conclusion by introducing a family of quantum communication protocols within the classical network framework and systematically evaluating their performance on both real and synthetic networks. Our claim is further supported by a theoretical analysis valid in the thermodynamic limit of random scale-free graphs.}
In this paper, \newchange{we address this question by showing that quantum
communication can be sustained on ultra-large-scale networks with
degree-heterogeneous and small-world structure, such as that observed in the
Internet~\cite{faloutsos1999power, watts1998collective,
barabasi1999emergence}.}
\newchange{We reach this conclusion by introducing a family of quantum
communication protocols within the classical network framework and
systematically evaluating their performance on both real network topologies
and synthetic networks. Our results are further supported by a theoretical
analysis in the thermodynamic limit of random scale-free graphs.}

\section{Results}

\subsection{Quantum-network representation and communication protocols}

%%%%%%%%%%%%%
\begin{figure*}[!htb]
    \centering
    \includegraphics[width=0.75\linewidth]{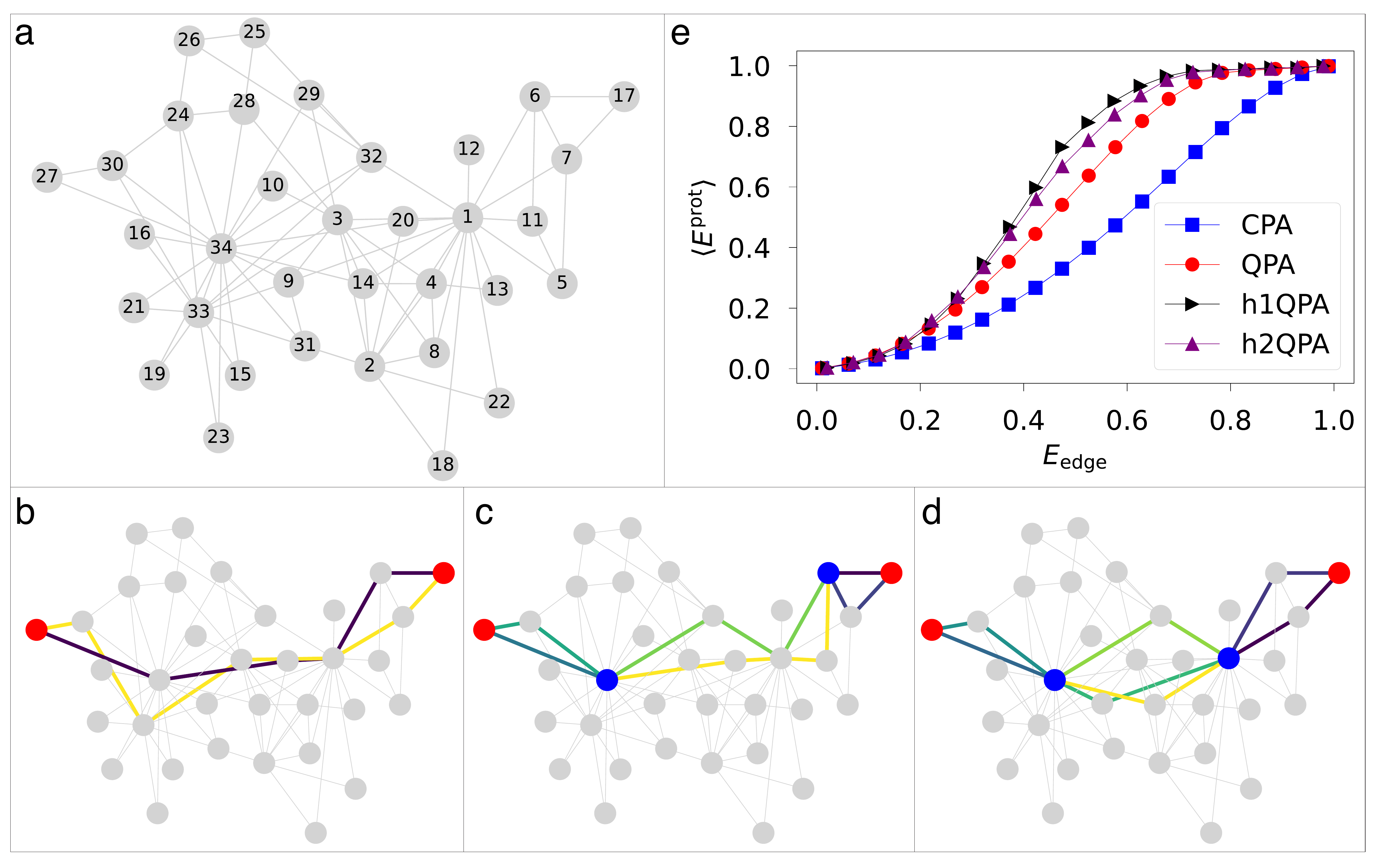}
    \caption{Quantum communication on a real network. (a) As an illustrative example, we consider the Zachary Karate Club's network~\cite{zachary1977information}. (b) We set the entanglement of the individual edges to $E_\textrm{edge} = 0.60$, then we apply the QPA protocol to establish a communication channel between the nodes $s=17$ and $t=27$, both highlighted in red. Edges utilized by the protocol are visualized with a color different from gray. In particular, all edges belonging to the same path selected by the QPA protocol have the same color. The resulting entanglement is $E^\textrm{QPA}_{s, t} = 0.41$. In this case, the CPA protocol selects the very same paths as QPA, however, the resulting entanglement is $E^\textrm{CPA}_{s, t} = 0.12$. (c) Same as in (b), but for the h1QPA protocol. Here, we denote with blue circles the repeaters $r_s = 6$ and $r_t = 34$. We obtain $E^\textrm{h1QPA}_{s, t} = 0.59$. (d) Same as in (b) and (c), but for the h2QPA protocol. The repeaters are $r_s = 1$ and $r_t = 34$ and the resulting entanglement is $E^\textrm{h2QPA}_{s, t} = 0.79$. \change{(e)} We plot the value of the entanglement $\langle E^{\textrm{prot}} \rangle$ averaged over $P=10,000$ randomly selected pairs of nodes $s$ and $t$ as a function of the entanglement of the individual edges, see Eq.~(\ref{eq:average_performance}). Different curves correspond to different communication protocols. As overall metric of performance of a protocol, we measure the area under the curve (AUC), see Eq.~(\ref{eq:auc}), finding that AUC$^{\textrm{h2QPA}} =  0.60$, 
AUC$^{\textrm{h1QPA}} =   0.61$, AUC$^{\textrm{QPA}} = 0.55$, 
and AUC$^{\textrm{CPA}} =   0.42$.
}
    \label{fig:1}
\end{figure*}
%%%%%%%%%%%%%%%%%%%%%

The starting point of our study is the  mapping proposed by Acin {\it et al.} according to which a quantum network can be seen as a classical weighted graph~\cite{acin2007entanglement}. The presence of the edge $(i,j)$ in the classical graph indicates that the state $\ket{\lambda_{i,j}}$ is composed of partially entangled qubits with Schmidt coefficient  $\lambda_{i,j} \in [1/2, 1]$, i.e., Eq.~(\ref{eq:qubit}). The weight of the edge $(i,j)$ is equal to the entanglement of the state $\ket{\lambda_{i,j}}$, i.e., $ E(\ket{\lambda_{i,j}}) = 2 (1 - \lambda_{i,j})$, see Eq.~(\ref{eq:ent}). 
\newchange{Thus, the framework explicitly accounts for the quality of
the initial entanglement associated with individual network links, while
abstracting away the physical processes through which such entanglement
is generated and maintained.}
The mapping allows us to treat the operations of entanglement swapping and distillation as  transformations of the edges of the classical graph, see~\EM~\ref{app:A} for details.

%Each member of the proposed family of communication protocols operates on a  weighted annealed graph to establish practical rules for the creation of a quantum communication channel between an arbitrary pair of nodes $s$ and $t$. 
Each member of the proposed family of communication
protocols operates on a weighted annealed graph
\newchange{to create a quantum communication channel}
between an arbitrary pair of nodes $s$ and $t$.
The common components of all methods within the family are: 
(i) A greedy selection of optimal paths between suitably chosen pairs of nodes in the graph; (ii) The application of the operation of entanglement swapping, i.e., Eq.~(\ref{eq:multi_swapping}), to the  paths selected at point (i) to create new states between far-apart nodes; (iii) The application of the operation of entanglement distillation, i.e., Eq.~(\ref{eq:multi_distil}), to the  states constructed at point (ii) to enhance the quality of the overall quantum communication channel.

The various protocols within the family differ in how the communication channel between nodes $s$ and $t$ is actually constructed. In the \change{Quantum Path Aggregation (QPA)} protocol, individual paths between nodes $s$ and $t$ are formed with the objective of maximizing the entanglement obtained via swapping.  Classically, the protocol can be seen as a greedy algorithm for the solution of the so-called disjoint-edge shortest path problem~\cite{eilam1998disjoint}, which is relevant for various routing applications~\cite{bhandari1999survivable}.
  Also, QPA can be seen as a non-trivial generalization of the protocol proposed by Malik {\it et al.} to approximate so-called $st$-concurrence percolation~\cite{malik2022concurrence}. 
  Their approximation consists in considering all paths as self-avoiding and of identical length. Malik {\it et al.}'s  method is conceived for networks with identical weights; it was applied to graphs with size up to $N = 10^4$. Building on prior work on path percolation~\cite{kim2024shortest, kim2025modeling, kim2026shortest}, our proposed protocol does not only overcome the above limitations and approximations, but can also be efficiently applied to much larger networks.
  \change{Path aggregation protocols have been used in quantum communication~\cite{azuma2017aggregating, lo2020aggregating}. Compared to link-level protocols, they operate at a global scale by establishing independent end-to-end paths. From a hardware implementation standpoint, this is significantly less demanding because quantum
memory is required only at the two end nodes, but not at intermediate nodes along the path.}

We generalize QPA into a heterogeneous version, which we denote as hxQPA protocol, with $x =0, 1, 2, \ldots$ In hxQPA, two high-degree nodes $r_s$ and $r_t$ in the neighborhoods of $s$ and $t$, respectively, are appropriately selected to segment the communication channel $s \to t$ into the three sub-channels $s \to r_s$, $r_s \to r_t$ and $r_t \to t$. The rationale is that the channels $s \to r_s$ and $r_t \to t$ sustain communication between nearby nodes, whereas the channel $r_s \to r_t$ serves a potentially long-range communication.
\newchange{By selecting high-degree nodes as repeaters, this construction
exploits degree heterogeneity to increase the redundancy of the long-range
communication channel, with the goal of compensating for the growth of
communication distances with network size.}
The parameter $x$ denotes the radius of the neighborhoods of the nodes $s$ and $t$, tuning therefore the actual range of the short-distance sub-channels. In particular, for $x=0$, the heterogeneous QPA protocol is the same as  QPA.

Also, we consider the protocol
\change{Classical Path Aggregation (CPA).}
%Classical Entanglement Percolation (CPA), where no operations of entanglement swapping and distillation are used. Quantum communication is instead seen as a purely 
\change{In that, quantum communication is seen as a}
classical percolation process in the graph, where each edge $(i,j)$ is considered occupied with probability equal to $E(\ket{\lambda_{i,j}})$. Percolation paths are still greedily selected with the goal of maximizing the probability to percolate from node $s$ to $t$, i.e., Eq.~(\ref{eq:best_path_CEP}). Then, the overall probability to percolate from $s$ to $t$ is computed as the probability that at least one of the optimal paths percolates, i.e., Eq.~(\ref{eq:CEP_protocol}).
From the practical point of view, the CPA protocol uses the same rules as QPA, but optimizes a different objective function, see \EM~\ref{app:A}. 
Also, the entanglement created by CPA is always bounded from above by the one created by QPA.

%We stress that, although the expressions CPA and QPA are taken from past literature~\cite{acin2007entanglement}, the proposed protocols have been not considered previously.

In Fig.~\ref{fig:1}, we provide an exemplificative  application of the various protocols to the Zachary Karate Club's network~\cite{zachary1977information}. 
%We recognize this is a social network, thus not particularly relevant in the study of quantum communication; on the other hand, this network dataset is so iconic that is ideal for illustrative purposes. In our example of the application of the communication protocols, 
Here and in the rest of the paper, we start from an initially unweighed network, then we add identical weights $0 \leq E_\textrm{edge} \leq 1$ to all edges while studying the performance of the quantum communication protocols.
%We assume that the weight of each edge in the network is the same, i.e., $E(\ket{\lambda_{i,j}}) = E_\textrm{edge}$ for all edges $(i,j)$. 
Figs.~\ref{fig:1}b, c and d respectively display the paths identified by the QPA, h1QPA and h2QPA protocols. In this specific case, the paths found by CPA are the same as those found by QPA. We stress that in QPA, no repeaters are used; also, the repeaters used in h1QPA are different from those used in h2QPA. The illustrative examples of Figs.~\ref{fig:1}b, c and d are valid for a specific value of $ E_\textrm{edge}$, and for a specific pair of nodes $s$ and $t$. To get an overall estimate of the performance of the generic protocol ``$\textrm{prot}$,'' we compute the average value of the entanglement over many randomly selected pairs of nodes $s$ and $t$, denoted as $\langle E^{\textrm{prot}} \rangle$, as $ E_\textrm{edge}$ is varied, see Eq.~(\ref{eq:average_performance}).
As shown in Fig.~\ref{fig:1}e, there is a large variability of performance between the various protocols. For example, CPA is never able to generate perfect quantum communication channels in a network that is composed of non-maximally entangled states, i.e., $\langle E^{\textrm{CPA}} \rangle < 1 $ as long as $E_\textrm{edge}  < 1$; for QPA, we have $\langle E^\textrm{QPA} \rangle = 1 $ for $E_\textrm{edge}  > 0.8$, denoting that maximum entanglement can be reached even in networks that are not formed by maximally entangled states; the performance of the h1QPA and h2QPA protocols reaches such a saturation point for even smaller values of  $E_\textrm{edge}$. As a simple and standardized metric of performance, we compute the area under the curves (AUCs) of Fig.~\ref{fig:1}e, i.e., Eq.~(\ref{eq:auc}). In the specific example of the Zachary Karate Club's network, we find that h2QPA and h1QPA have nearly identical performance, followed by QPA and CPA.

\subsection{Quantum communication on synthetic networks
and large-scale behavior of path aggregation}

\begin{figure}[!htb]
\centering
\includegraphics[width=0.45\textwidth]{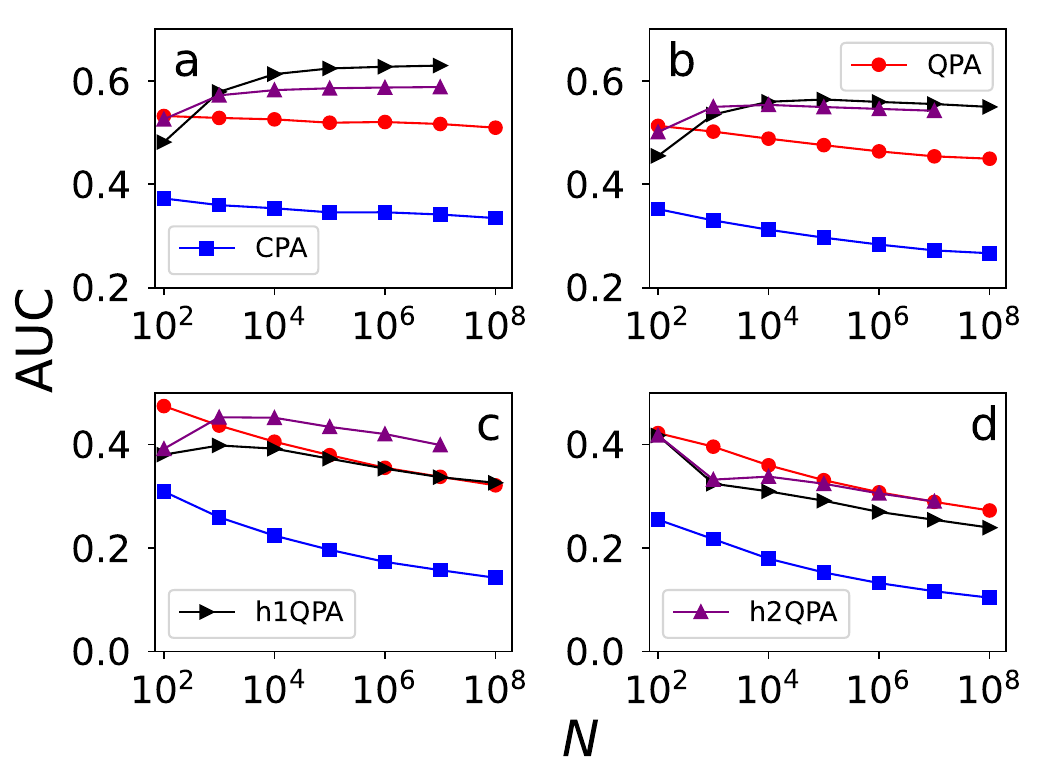}
\caption{Quantum communication on synthetic networks. (a) We measure the Area Under the Curve [AUC, see Eq.~(\ref{eq:auc})] for the various communication protocols on random graphs with power-law degree distribution $\mathcal{K}^{(0)}(k) \sim k^{-\gamma}$. The AUC metric is plotted as a function of the network size $N$ for $\gamma = 2.1$. Results are averaged over multiple instances of the network model, and many randomly selected pairs of nodes $s$ and $t$, see Appendix~\ref{app:configuration_model} for details. (b-d) Same as in (a), but for $\gamma = 2.5, 3.5$ and $4.5$, respectively.}
\label{fig:2}
\end{figure}

Next, we systematically study the performance of the various protocols on instances of the uncorrelated configuration model with power-law degree distribution $\mathcal{K}^{(0)}(k) \sim k^{-\gamma}$, with $\gamma > 2$, see Appendix~\ref{app:configuration_model} for details~\cite{molloy1995critical, catanzaro2005generation}.
As the results of Fig.~\ref{fig:2} show, the performance of both the QPA and CPA protocols decreases as $N$ increases for any value of the degree exponent $\gamma > 2$. The finding is explained by the following argument. The best quantum communication channel that can be obtained via QPA is from the distillation of $k$ paths of length $\ell$. Here, $k =  \min \{k_s, k_t\}$ indicates the minimum value of the degree of the nodes $s$ and $t$, whereas $\ell$ is their geodesic distance. We refer to $k$ as the redundancy of the communication channel, and to $\ell$ as its length.
For given $k$ and $\ell$, the minimum value of the 
individual edge entanglement $E^*_\textrm{edge}( k, \ell; E_{\textrm {target}})$
required to achieve a 
%perfect 
quantum communication channel 
with desired entanglement $E_{\textrm {target}} \in (0, 1]$ is given by 
Eq.~(\ref{eq:partial_entanglement}).
For a perfect communication channel $E_{\textrm {target}} = 1$, it simplifies to
\begin{equation}
E^*_\textrm{edge}( k, \ell; E_{\textrm {target}}=1) = 1 - \sqrt{ 1 - 2^{\frac{2k-1}{k \, \ell}} \left(  1 - 2^{-\frac{1}{k}} \right)^{\frac{1}{\ell}} } \, .
\label{eq:min_entanglement}
\end{equation}
The limiting behavior of this quantity depends on how $k$ and $\ell$ behave as the system size $N$ increases, see Eq.~(\ref{eq:main_limit_results}). 
For QPA, the probability that $k$ is finite approaches one as $N$ increases. However, $\ell$ diverges with $N$. Specifically, for $2 < \gamma \leq 3$, the configuration model displays ultra-small-world behavior, i.e.,  $\ell \sim \log \log N$~\cite{cohen2003scale}; for $\gamma > 3$, the typical distance of the nodes in the graph is instead $\ell \sim \log N$, as expected for small-world networks~\cite{watts1998collective}. 
In summary, $\lim_{N \to \infty} E^*_\textrm{edge}( k \sim \textrm{const.}, \ell; E_{\textrm {target}}=1 ) = \lim_{\ell \to \infty} E^*_\textrm{edge}( k \sim \textrm{const.}, \ell; E_{\textrm {target}}=1 ) = 1 $. Such a limit is valid not just for a maximally entangled channel but also for any non-null level of desired entanglement $E_{\textrm {target}}$, thus $\lim_{N \to \infty} \langle E^\mathrm{QPA} \rangle = 0$ unless $E_\textrm{edge} = 1$, leading therefore to $\lim_{N \to \infty}$ AUC$^{\mathrm{QPA}} = 0$ for any value of the degree exponent $\gamma > 2$, see Appendix~\ref{app:thermodynamic}.  This means that, in the thermodynamic limit, the QPA protocol is not sustainable as it can not allow communication between a sizable number of random pairs of nodes unless the network is characterized by maximally entangled states. We stress that such a thermodynamic behavior is approached extremely slowly for scale-free graphs with $2 < \gamma \leq 3$, as demonstrated from the fact that the decrease in the AUC metric from $N=10^2$ to $N=10^8$ is barely noticeable, see Fig.~\ref{fig:2}a and b. 
%From practical purposes therefore, the protocol can be still useful. 
Since the performance of CPA is bounded from above by QPA, the CPA protocol is also not sustainable in the thermodynamic limit.

%%%%%%%%%%%%
%%%%%%%%%%%%

The issue faced by QPA in the limit of infinitely large networks is due to the contrast between the finite  redundancy $k$ of the channel that can be constructed between two randomly chosen nodes, and the divergence of the length $\ell$ of the channel. The hxQPA protocols are specifically designed to avoid such a problem by segmenting the communication channel $s \to t$ into the three sub-channels $s \to r_s$, $r_s \to r_t$ and $r_t \to t$. The sub-channel $r_s \to r_t$ is devoted to long-range communication. This is still constructed according to the rules of the standard QPA protocol, however, the repeaters $r_s$ and $r_t$ have degrees $k_{r_s}$ and $k_{r_t}$, respectively, whose values grow sufficiently fast to compensate the divergence of  $\ell$. At the same time, the short-range portions of the channel, namely $s \to r_s$ and $r_t \to t$, involve communication between nodes that are at most distance $x$, regardless of the network size. The fact that the degrees $k_{r_s}$ and $k_{r_t}$ can assume potentially very large values is due to a principle known in the literature of network theory as the friendship paradox~\cite{feld1991your}. Mathematically, this can be understood using the degree distribution $\mathcal{Q}(k) \sim k^{1 - \gamma}$ of a node found at the end of a randomly chosen edge, i.e., the so-called excess degree distribution~\cite{newman2018networks}. As shown in Appendix~\ref{app:redundancy}, the degrees $k_{r_s}$ and $k_{r_t}$ can be estimated using extreme value theory applied to $\mathcal{Q}(k)$. At the same time, the size of the sample used to estimate the extreme value is determined by the radius $x$ of the node's neighborhood. For $x < 2$, the neighborhood is not large enough to compensate for the decay of $\mathcal{Q}(k)$, thus the resulting protocol is not sustainable in the thermodynamic limit. However, for $x \geq 2$ and $2 < \gamma < 3$, the size of the neighborhood explodes with growing network size~\cite{newman2018networks}, and the probability that the variable $k = \min\{ k_{r_s}, k_{r_t}\}$ grows at least as $N^\alpha$ does not vanish as $N$ increases, for any $0 < \alpha < \frac{3-\gamma}{2}$ (i.e., as long as $N^\alpha$ grows to infinity slower than the second moment of the degree distribution $\langle k^2 \rangle \sim N^{(3-\gamma)/2}$), see Eq.~(\ref{eq:main_limit_results2}). For $\gamma=3$, instead, the probability that $k$ grows at least as $(\log N)^\alpha$ does not vanish for any $0<\alpha<1$, see Appendix~\ref{app:thermodynamic}. One can then prove, see Eq.~(\ref{eq:main_limit_results}), that 
%the asymptotic minimum value of the entanglement required to achieve perfect communication in the long-range portion of the channel, i.e., Eq.~(\ref{eq:min_entanglement}), is
the limiting behavior of Eq.~(\ref{eq:min_entanglement}) is
\begin{equation}
\lim_{N \to \infty}  E^*_{\textrm{edge}}( k \sim N^\alpha, \ell \sim \log N; E_{\textrm {target}}=1) = 1 - \sqrt{1 - e^{-\alpha}}
\; .
\label{eq:lim_min_ent}
\end{equation}
This indicates that the sub-channel $r_s \to r_t$ can sustain perfect communication also in an infinitely large scale-free network even if the network is not maximally entangled, provided that the network is at least in the small-world regime. In the ultra-small-world regime, we have that
the very same limit of Eq.~(\ref{eq:lim_min_ent}) holds even for the much slower scaling $k \sim \left(\log N \right)^\alpha$. Also, we can prove that $\lim_{N \to \infty}  E^*_{\textrm{edge}}( k \sim N^\alpha, \ell \sim \log \log N; E_{\textrm {target}}=1) = 0$
\newchange{meaning that, for $2 < \gamma < 3$, sufficiently
rapid growth of channel redundancy allows network-wide quantum communication
for arbitrarily small nonzero values of the initial edge entanglement.
The above limits are reported for the special case $E_{\textrm {target}}=1$, however, they hold for arbitrary $0 < E_{\textrm {target}} \leq 1$.}
%, meaning that, if channel redundancy grows quickly enough, then network-wide quantum communication is feasible for any non-null level of initial entanglement.

The above argument does not exclude the potential degradation of communication in the short-range sub-channels. This is, however, not expected to occur as scale-free graphs are rich in short loops~\cite{bianconi2005loops}. In addition, 
our findings are
confirmed in networks with non-vanishing clustering coefficient constructed according to the $\mathbb{S}^1$ model, see Fig.~\ref{fig:SMs1}~\cite{serrano2008self}. 
Once more, we stress that the thermodynamic limit is reached very slowly. For $2 < \gamma \leq 3$, even networks with $N = 10^8$ do not display an apparent degradation in h1QPA performance. Our results on large, but finite networks show that h1QPA outperforms h2QPA; our theory predicts, however, that h1QPA performance will eventually drop to zero as the network size increases, whereas the  h2QPA performance will not vanish.

\subsection{Quantum communication on real networks}

%%%%%%%%%%%%
\begin{figure}[!htb]
\centering
\includegraphics[width=0.45\textwidth]{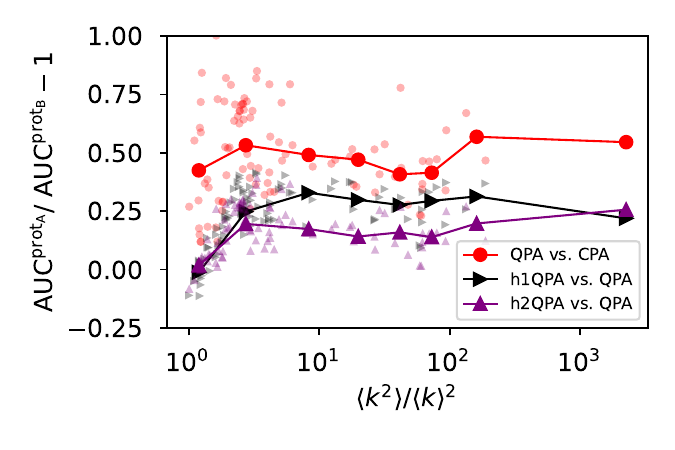}
\caption{
Quantum communication on real networks.
We display with small red circles the  relative improvement AUC$^{\mathrm{QPA}}$/AUC$^{\mathrm{CPA}} - 1$ of QPA {\it vs.} CPA for 
$101$ real networks. 
Each network is further characterized by a different level of degree heterogeneity $\langle k^2 \rangle / \langle k \rangle^2$ used as abscissa in the plot. 
To highlight the trend, we group data using logarithmic binning and display with large red circles  median values for each of the bins. Black and purple triangles display analogous results, but for the comparisons h1QPA {\it vs.} QPA and h2QPA {\it vs.} QPA, respectively.
}
\label{fig:3}
\end{figure}
%%%%%%%%%%%%

%Real graphs
We conclude our paper by reporting on a systematic analysis performed on a corpus composed of $101$ real networks, see Fig.~\ref{fig:3}. The corpus contains rather different networks, e.g.,  biological, social and technological networks; networks in the corpus also display a large variability in size and topological properties, as for example degree distribution, average distance, and clustering coefficient. Details on the networks as well as the corresponding AUC values are reported in Tables~\ref{tab:real1} and~\ref{tab:real2}. The conclusions drawn for synthetic graphs are confirmed on real networks. The quantum advantage in using QPA instead of CPA is roughly the same for all networks, irrespective of their size and properties. The heterogeneous variants of the QPA protocol outperform standard QPA, but only if networks are sufficiently large and heterogeneous. The h1QPA and h2QPA protocols display comparable performance, with a slight advantage of h1QPA due to the finite size of the networks.

\section{Conclusions}

On par with  previous literature, this paper tacitly assumes that each individual node has global knowledge of the network structure stored in a potentially very large routing table. Also, the protocols are devised for the establishment of communication of one pair of nodes at a time. Once a pair of nodes has communicated using some of the network's resources, then these resources are fully restored before another pair of nodes attempts to communicate. \change{From this perspective, characterizing the critical properties of the proposed protocols under the assumption of finite resources could shed additional light on their performance in more realistic settings~\cite{kim2024shortest, kim2025modeling, kim2026shortest}.} 
We believe that future work should be devoted to addressing these limitations.
\newchange{Another direction for future work is to determine
whether the topology-driven scaling behavior we identified persists
when dynamical noise, decoherence, finite memory times, and other
implementation-dependent imperfections are explicitly incorporated.}
\newchange{Our results establish that, within the network representation
considered here, sustainable quantum communication is controlled by the
competition between communication distance and path redundancy. Degree
heterogeneity can provide increasing redundancy that compensates for
growing communication distances in small-world networks, thereby enabling
quantum communication at ultra-large scales. Scale-free and small-world
structures are widespread in real complex networks, including the
Internet~\cite{barabasi1999emergence,pastor2004evolution}, suggesting that
network topology may play an important role in determining the scalability
of future quantum communication networks.}

\acknowledgments{
This research was partially supported by the Air Force Office of Scientific
Research under grant number FA9550-24-1-0039. The funders had no role
in study design, data collection, and analysis, the decision to publish, or
any opinions, findings, conclusions, or recommendations expressed in
the manuscript. The author used ChatGPT (OpenAI) to assist with language editing, presentation, and aspects 
of the scientific framing of the manuscript. 
All scientific content, analyses, and conclusions were independently reviewed and verified by the author.
}

%The 
\section*{Data availability}
Code implementing the protocols introduced in this paper is available at \url{https://github.com/filrad/QuantumCommunicationProtocols}.
All network datasets considered in the paper are publicly available. Data-source references are provided in Tables~\ref{tab:real1} and~\ref{tab:real2}.

% \EM~
%\section*{End Matter}

\appendix

\section{Quantum networks and protocols of communication}
\label{app:A}

%\subsection{Quantum networks and protocols of communication}

In the classical representation of a quantum network~\cite{acin2007entanglement}, an edge between two nodes $i$ and $j$ stands for a pair of entangled qubits that in the Schmidt form is written as
\begin{equation}
\ket{\lambda_{i,j}} = \sqrt{\lambda_{i,j}} \ket{00} + \sqrt{1 - \lambda_{i,j}} \ket{11} \; ,
\label{eq:qubit}
\end{equation}
with Schmidt coefficient $\lambda_{i,j} \in \left[\frac{1}{2}, 1 \right]$. Not all pairs of nodes are connected in the graph by an edge. In particular, the degree $k_i$ represents the number of connections that node $i$ has with  other nodes in the graph.
The weight of the edge $(i,j)$ is given by the entanglement of the state $\ket{\lambda_{i,j}}$ defined as 
\begin{equation}
E(\ket{\lambda_{i,j}}) = 2 (1 - \lambda_{i,j} ) \; ,
\label{eq:ent}
\end{equation}
i.e., the probability to convert $\ket{\lambda_{i,j}}$ to a singlet or Bell state~\cite{acin2007entanglement}. Note that the graph representation admits multi or parallel edges between the same pair of nodes $i$ and $j$. 

Two main operations of entanglement distribution can be performed to create entanglement between arbitrary pairs of nodes $s$ and $t$. The operation of entanglement swapping is applied to the edges in the path $\mathbf{p}_{s, t} = (s=i_0, i_1, \ldots, i_\ell = t)$ to create a new state whose Schmidt coefficient equals
\begin{equation}
^{(\mathrm{swa})}\lambda_{s,t}  = \frac{1 + \sqrt{ 1 - 2^{2 \ell} \, \prod_{i=1}^{\ell} \lambda_{i-1,i} (1 - \lambda_{i-1,i})} }{2} \;. 
\label{eq:multi_swapping}
\end{equation}
The operation of entanglement distillation is applied to the parallel edges 
$\ket{\lambda^{(1)}_{s,t}}$, $\ket{\lambda_{s,t}^{(2)}}$, $\ldots$, $\ket{\lambda_{s,t}^{(k)}}$ to create a new state with Schmidt coefficient equal to
\begin{equation}
^{(\mathrm{dis})}\lambda_{s, t} = \max \left\{ \frac{1}{2}, \prod_{i=1}^k \, \lambda_{s, t}^{(i)} \right\} \; . 
\label{eq:multi_distil}
\end{equation}

 \subsection{Quantum Path Aggregation}
 %{\it QPA protocol.} 
 The aim of the
 protocol is to create a pair of entangled qubits between two potentially far away nodes $s$ and $t$. The protocol consists of entanglement swapping operations along optimal paths between $s$ and $t$, followed by the distillation of the resulting states. Specifically, we set $k = 0$ and iterate the following:

\begin{enumerate} 

\item Identify all paths existing in the network between $s$ and $t$. If no path is identified, exit from the iterative algorithm. Otherwise, increase the value of the variable $k \mapsto k + 1$ and consider the optimal path (breaking ties at random), i.e.,
\begin{equation}
\mathbf{\hat{p}}^{(k)}_{s, t} = \arg \, \change{\min_{\mathbf{p}_{s, t}}} \;  \;^{\textrm{(swa)}}\lambda(\mathbf{p}_{s, t})  \; .
\label{eq:best_path}
\end{equation}

\item  All edges of the path $\mathbf{\hat{p}}^{(k)}_{s, t}$ are used in a swapping operation and are therefore deleted from the network. In turn, the new state $\ket{\mu_{s, t}^{(k)}}$ is created.  Note that the state $\ket{\mu_{s, t}^{(k)}}$ is not used when searching for additional optimal paths at point 1.

\item Estimate the entanglement between nodes $s$ and $t$ at stage $k$ of the algorithm as
\begin{equation}
E_{s, t}^{(k)}   = 
2 \left( 1 - \max{ \left\{ \frac{1}{2} , \prod_{z =1}^k \mu_{s , t}^{(z)} \right\}} \right) \; .
\label{eq:dist_protocol}
\end{equation}
%This is nothing more than the entanglement of Eq.~(\ref{eq:ent}) obtained by applying the distillation  rule of Eq.~(\ref{eq:multi_distil}) to the states $\ket{\mu_{s\to t}^{(0)}}, \ket{\mu_{s, t}^{(1)}}, \ldots, \ket{\mu_{s\to t}^{(k)}}$. 
If $E_{s, t}^{(k)} = 1$,  exit from the iterative algorithm. Otherwise,  go back to point 1.

\end{enumerate}

At the end of the algorithm, $E_{s, t}^{(k)} = E_{s, t}^{\mathrm{QPA}}$ quantifies the entanglement between nodes $s$ and $t$. 
\newchange{The objective of QPA is therefore to maximize the entanglement
generated between the end nodes, which differs from approaches that
minimize the resource cost required to satisfy a prescribed
entanglement-distribution task~\cite{azuma2025networking}.}

  \subsection{Classical Path Aggregation}
 The CPA protocol  relies on the same steps as above. However, the decision step of Eq.~(\ref{eq:best_path}) is replaced by
\begin{equation}
\mathbf{\hat{p}}^{(v)}_{s, t} = \arg \, \max_{\mathbf{p}_{s, t}} \; \;  \; \prod_{i=1}^\ell 2 (1 - \lambda_{i-1, i}) =  \arg \, \max_{\mathbf{p}_{s, t}} P^\textrm{CPA}(\mathbf{p}_{s, t})\;  \; ,
\label{eq:best_path_CEP}
\end{equation}
where the objective function $P^\textrm{CPA}(\mathbf{p}_{s, t})$ represents the  probability of
percolating along the path.
Entanglement at stage $k$ of the protocol is measured as
\begin{equation}
\tilde{E}_{s, t}^{(k)}   = 1 - \prod_{z=1}^k
 \left[ 1 - P^\textrm{CPA}(\mathbf{\hat{p}}^{(z)}_{s, t}) \right] \; ,
\label{eq:CEP_protocol}
\end{equation}
i.e., the probability to percolate between $s$ and $t$ along at least one of the identified optimal paths. At the end of the algorithm, denote with $E^{\textrm{CPA}}_{s,t}$ the entanglement of the quantum channel generated by the CPA protocol.
The quantum operations used by QPA bound the corresponding CPA probabilities from above: swapping along any path gives $E(\ket{^{(\mathrm{swa})}\lambda_{s,t}}) \geq \prod_{j=1}^{\ell} E(\ket{\lambda_{j-1,j}})$, while distillation of parallel paths gives $E(\ket{^{(\mathrm{dis})}\lambda_{s,t}}) \geq 1-\prod_j[1-E(\ket{\lambda_{s,t}^{(j)}})]$. Thus, $E^{\textrm{QPA}}_{s,t} \geq E^{\textrm{CPA}}_{s,t}$.

 \subsection{Heterogeneous Quantum Path Aggregation}

 The h$x$QPA protocol generalizes QPA by breaking the communication channel between $s$ and $t$ into up to three sub-channels disposed in series. 
Denote with $\mathcal{N}_s(x)$ the set of all nodes at distance at most $x = 0, 1, 2, \ldots$ from node $s$, with node $s$ included. 
%Note that $s \in \mathcal{N}_s(x)$ for all $x \geq 0$. 
In the h$x$QPA protocol, one first identifies 
the repeater $r_s$ of node $s$ as either $t$ or the largest-degree node in $\mathcal{N}_s(x)$, i.e., 
$r_s = 
\arg \max_{q \in \mathcal{N}_s(x) } k_q$ if $t \notin \mathcal{N}_s(x)$ or $r_s = t$ otherwise. A similar definition applies to the repeater $r_t$ of node $t$. Then, the standard QPA protocol is used to establish the quantum communication channels 
$s \to r_s$, $t \to r_t$ and $r_s \to r_t$. Note that one network edge can be used only once, 
thus the three channels rely on disjoint sets of edges.
%Specifically, the order of construction of the channels is $s \to r_s$, $r_t \to t$, and $r_s \to r_t$. 
The three channels have associated the coefficients $\mu^{\textrm{QPA}}_{s , r_s} = 1 - E^{\textrm{QPA}}_{s , r_s}/2$, 
$\mu^{\textrm{QPA}}_{t , r_t} = 1 - E^{\textrm{QPA}}_{t , r_t}/2$ and $\mu^{\textrm{QPA}}_{r_s \to r_t} = 1 - E^{\textrm{QPA}}_{r_s , r_t}/2$, respectively.
These three channels are arranged in series, thus 
entanglement swapping is used to 
%one needs to apply the rule of  
%entanglement swapping to 
generate the state with coefficient 
\begin{equation}
\begin{array}{l}
\mu^{\textrm{hxQPA}}_{s , t} = \frac{1}{2} + \frac{1}{2} \left[ 1 - 64 \; \mu^{\textrm{QPA}}_{s , r_s} (1- \mu^{\textrm{QPA}}_{s , r_s})  \right.
\times
\\
\times \left. \mu^{\textrm{QPA}}_{r_s , r_t} (1- \mu^{\textrm{QPA}}_{r_s , r_t}) \times \mu^{\textrm{QPA}}_{t , r_t} (1- \mu^{\textrm{QPA}}_{t , r_t}) \right]^{1/2} 
\end{array} \; .
\label{eq:hQPA_channel}
\end{equation}
The entanglement between nodes $s$ and $t$ is finally obtained using
Eq.~(\ref{eq:ent}), i.e., $E^{\mathrm{hxQPA}}_{s , t} = 2 \left( 1 - \mu^{\textrm{hxQPA}}_{s , t} \right)$.
%\begin{equation}
%E^{\mathrm{hxQPA}}_{s , t} = 2 \left( 1 - \mu^{\textrm{hxQPA}}_{s , t} \right) \; .
%    \label{eq:hQPA}
%\end{equation}
In some cases, one might find that $r_s = s$, $r_s = r_t$, $t = r_t$,$r_s = t$ and/or $r_t = s$, resulting in less than three sub-channels. The above formulation is still valid by applying entanglement swapping between the number of effective sub-channels obtained via the standard QPA protocol. Notice that the hxQPA protocol reduces to the standard QPA for $x =0$.  In such a case, one has that $r_s = s$ and $r_ t = t$.

\subsection{Performance of quantum communication protocols}

%{\it Performance of quantum communication protocols.} 
To estimate the overall performance of the protocol ``$\textrm{prot}$'' over a network, one first obtains the average  $ \langle E^{\textrm{prot}} \rangle$ over the set $\mathcal{P}$ of size $P$ composed of randomly selected pairs of nodes $s$ and $t$ for a fixed value of the individual-edge entanglement $E_{\textrm{edge}}$, i.e., 
\begin{equation}
    \langle E^{\textrm{prot}} \rangle = \frac{1}{P} \sum_{s,t \in \mathcal{P}} E^{\textrm{prot}}_{s,t} \; ,
    \label{eq:average_performance}
\end{equation}

Then, one computes the area under the curve defined as
\begin{equation}
    \textrm{AUC} = \int_0^1 \, d E_{\textrm{edge}} \, \; \langle E^{\textrm{prot}} \rangle \; .
    \label{eq:auc}
\end{equation}
For all results reported in this paper, the above integral is approximated using $20$ distinct values of 
$E_{\textrm{edge}}$ equally spaced in the interval $[0, 1]$. The specific number of pairs $P$ used in the estimate of $ \langle E^{\textrm{prot}} \rangle$ varies from network to network, and from protocol to protocol, see Appendices~\ref{app:configuration_model}, \ref{app:s1}, and~\ref{app:real_networks} for details.

\section{Theoretical analysis}
\label{app:theory}

\begin{comment}
\subsection{Condition for perfect communication}

Given two nodes $s$ and $t$ with degrees $k_s$ and $k_t$, and
geodesic distance $\ell$, \newchange{we consider the ideal case in which
$k=\min\{k_s,k_t\}$ edge-disjoint paths of length $\ell$ are available
between them. The entanglement generated by QPA under these conditions
is given by Eq.~(1).}
This is obtained by setting $\lambda_{i,j} = \lambda$ in Eq.~(\ref{eq:multi_swapping}) to get $^{\textrm{(swa)}} \lambda_{s,t} = \frac{1}{2} \left[ 1  + \sqrt{1 - 2^{2\ell} \lambda^\ell (1 - \lambda)^\ell} \right] $. Then, one applies Eq.~(\ref{eq:multi_distil}) to $k$ of such states, thus the condition of a perfect channel is given by 
$\frac{1}{2^k} \left\{ 1  + \sqrt{1 - 2^{2\ell} \left[ \lambda^* (1 - \lambda^*) \right]^\ell} \right\}^k  = \frac{1}{2}$. From this, one gets
$\lambda^* = \frac{1}{2} \left[ 1 + \sqrt{ 1 - 2^{(2k-1)/(kl)} \left( 1 - 2^{-1/k}\right)^{1/\ell}} \right]$. Plugging this condition into Eq.~(\ref{eq:ent}) leads to Eq.~(\ref{eq:min_entanglement}).
\end{comment}

\newchange{

\subsection{Condition for target entanglement}

Given two nodes $s$ and $t$ with degrees $k_s$ and $k_t$, and
geodesic distance $\ell$, \newchange{we consider the ideal case in which
$k=\min\{k_s,k_t\}$ edge-disjoint paths of length $\ell$ are available
between them. We determine the minimum individual-edge entanglement
required to generate a communication channel with target entanglement
$E_{\rm target}\in(0,1]$.}
This is obtained by setting $\lambda_{i,j} = \lambda$ in
Eq.~(\ref{eq:multi_swapping}) to get
$^{\textrm{(swa)}} \lambda_{s,t} = \frac{1}{2} \left[ 1  + \sqrt{1 - 2^{2\ell} \lambda^\ell (1 - \lambda)^\ell} \right] $.
Then, one applies Eq.~(\ref{eq:multi_distil}) to $k$ of such states.
\newchange{Since a channel with entanglement $E_{\rm target}$ has
Schmidt coefficient
$\lambda_{\rm target}=1-E_{\rm target}/2$, the threshold condition is}
\[
\frac{1}{2^k}
\left\{
1+\sqrt{1-2^{2\ell}
\left[\lambda^*(1-\lambda^*)\right]^\ell}
\right\}^k
=
1-\frac{E_{\rm target}}{2}.
\]
\newchange{Solving for $\lambda^*$ gives}
\[
\lambda^*
=
\frac{1}{2}
\left[
1+
\sqrt{
1-
\left\{
4\left(1-\frac{E_{\rm target}}{2}\right)^{1/k}
\left[
1-\left(1-\frac{E_{\rm target}}{2}\right)^{1/k}
\right]
\right\}^{1/\ell}
}
\right].
\]
\newchange{Finally, plugging this condition into
Eq.~(\ref{eq:ent}), we obtain}
\begin{equation}
\begin{array}{l}
E_{\rm edge}^*(k,\ell;E_{\rm target})
=
\\
1-
\sqrt{
1-
2^{2/\ell}
\left(1-\frac{E_{\rm target}}{2}\right)^{1/(k\ell)}
\left[
1-\left(1-\frac{E_{\rm target}}{2}\right)^{1/k}
\right]^{1/\ell}
}.
\end{array}
\label{eq:partial_entanglement}
\end{equation}
\newchange{For $E_{\rm target}=1$, the above expression reduces to
Eq.~(\ref{eq:min_entanglement}) for a perfect communication channel.}
}

\newchange{
\subsection{Redundancy of communication channels in the configuration model}
\label{app:redundancy}

Let $\mathcal{K}^{(0)}(k)$ denote the degree distribution of the configuration model and
$\mathcal{K}^{(0)}(\leq k)$ its cumulative distribution. For the standard QPA protocol, the redundancy of the communication channel between two randomly selected nodes $s$ and $t$ is bounded by $k=\min\{k_s,k_t\}$. Its cumulative distribution is therefore
%\begin{equation}
\[
\mathcal{K}^{(0)}_{\min}(\leq k)
=
1-\left[1-\mathcal{K}^{(0)}(\leq k)\right]^2
=
2\mathcal{K}^{(0)}(\leq k)
-\left[\mathcal{K}^{(0)}(\leq k)\right]^2 .
%\label{eq:k0min}
%\end{equation}
\]

For the heterogeneous protocols, we need instead the degree statistics of nodes reached by exploring the neighborhoods of $s$ and $t$. The degree $q$ of a node reached by following a randomly selected edge obeys the cumulative distribution
%\begin{equation}
\[
\mathcal{Q}(\leq q)
=
\frac{1}{\langle k\rangle}
\sum_{k=k_m}^{q} k\,\mathcal{K}^{(0)}(k), 
%\label{eq:cdf_q}
%\end{equation}
\]
where $\langle k \rangle = \sum_k k \, \mathcal{K}^{(0)}(k)$ is the average degree.
For a node of degree $k$, the probability that all its neighbors have degree at most $q$ is consequently
\[
\mathcal{K}^{(1)}(\leq q|k)
=
\left[\mathcal{Q}(\leq q)\right]^k .
%\label{eq:k1_cond}
\]
Averaging over the degree of the initial node gives
\[
\mathcal{K}^{(1)}(\leq q)
=
\sum_k\mathcal{K}^{(0)}(k)
\left[\mathcal{Q}(\leq q)\right]^k
=
G_0\left[\mathcal{Q}(\leq q)\right],
%\label{eq:k1}
\]
where
\[
G_0(z)=\sum_k z^k\mathcal{K}^{(0)}(k)
\]
is the generating function of the degree distribution. Thus, for h1QPA, the cumulative distribution of the redundancy of the long-range channel is
%\begin{equation}
\[
\mathcal{K}^{(1)}_{\min}(\leq k)
=
2G_0\left[\mathcal{Q}(\leq k)\right]
-
\left\{
G_0\left[\mathcal{Q}(\leq k)\right]
\right\}^2 .
%\label{eq:k1_min}
\]
%\end{equation}

For h2QPA, consider a node of degree $k$ and denote by $v$ the number of edges leading from its neighbors to next-nearest neighbors. The excess degree associated with an edge obeys
%\begin{equation}
\[
\widetilde{\mathcal{Q}}(q)
=
\frac{(q+1)\mathcal{K}^{(0)}(q+1)}
{\langle k\rangle},
%\label{eq:excess}
%\end{equation}
\]
and $v$ is distributed according to the convolution
\begin{equation}
\mathcal{V}(v|k)
=
\sum_{q_1+\cdots+q_k=v}
\widetilde{\mathcal{Q}}(q_1)\cdots
\widetilde{\mathcal{Q}}(q_k).
\label{eq:convolution}
\end{equation}
It follows that the cumulative distribution of the largest degree among the second neighbors of a node of degree $k$ is
\[
\mathcal{K}^{(2)}(\leq q|k)
=
\sum_v
\mathcal{V}(v|k)
\left[\mathcal{Q}(\leq q)\right]^v ,
%\label{eq:k2_cond}
\]
and, after averaging over the degree of the initial node,
\[
\mathcal{K}^{(2)}(\leq q)
=
\sum_k
\mathcal{K}^{(0)}(k)
\mathcal{K}^{(2)}(\leq q|k).
%\label{eq:k2}
\]
The corresponding distribution of the redundancy of the long-range channel is
\[
\mathcal{K}^{(2)}_{\min}(\leq k)
=
2\mathcal{K}^{(2)}(\leq k)
-
\left[\mathcal{K}^{(2)}(\leq k)\right]^2 .
%\label{eq:k2_min_exact}
\]

The convolution in Eq.~(\ref{eq:convolution}) becomes difficult to evaluate for large networks. Its mean can be obtained using the generating function
$G_1(z)=G'_0(z)/G'_0(1)$ of the excess-degree distribution, yielding
\[
\langle v\rangle
=
k\,\frac{\langle k^2\rangle-\langle k\rangle}
{\langle k\rangle}
\equiv k S_2 ,
\qquad
S_2=
\frac{\langle k^2\rangle-\langle k\rangle}
{\langle k\rangle}.
%\label{eq:mean_second_neighbors}
\]
Approximating $\mathcal{V}(v|k)$ by a distribution concentrated around its mean therefore gives
\[
\mathcal{K}^{(2)}(\leq q|k)
\simeq
\left[\mathcal{Q}(\leq q)\right]^{kS_2},
\]
and hence
\[
\mathcal{K}^{(2)}_{\min}(\leq k)
\simeq
2G_0\left[\mathcal{Q}^{S_2}(\leq k)\right]
-
\left\{
G_0\left[\mathcal{Q}^{S_2}(\leq k)\right]
\right\}^2 .
%\label{eq:k2_min}
\]
We stress that this approximation is intended to capture the scaling behavior; the relative width of $\mathcal{V}(v|k)$ can grow with network size in strongly heterogeneous networks.
The same argument can be generalized to $x>2$ by replacing $S_2$ with the corresponding branching factor $S_x$, whose scaling is determined by successive powers of the excess-degree branching factor. 
}

\subsection{Thermodynamic behavior}
\label{app:thermodynamic}

%{\it Thermodynamic behavior.} 
The main results concerning the performance of the hxQPA protocol
on the configuration model in the thermodynamic limit are established by estimating
the limit
\begin{equation}
\lim_{N \to \infty} \left[ 1 - 2^{-\frac{1}{k(N)}} \right]^{\frac{1}{\ell(N)}} = \left\{
\begin{array}{ll}
e^{-\alpha} & \textrm{ if } \lim_{N \to \infty} \frac{\log k(N)}{\ell(N)} = \alpha
\\
0 & \textrm{ if } \lim_{N \to \infty} \frac{\log k(N)}{\ell(N)} =  \infty
\end{array}
\right.
 \; ,
    \label{eq:main_limit_results}
\end{equation}
where both $k(N)$ and $\ell(N)$ are non-negative functions of the network size $N$, and $\alpha \geq 0$. In the specific case of this paper, $\ell(N) = \log \log N$ for $ 2 < \gamma \leq 3$ or $\ell(N) = \log N$ for $\gamma > 3$;  $k(N)$ potentially diverges as $N$ grows.  The argument of the above limit is one of the two terms that depend on $k$ and $\ell$  in Eq.~(\ref{eq:min_entanglement}); for the other term, one trivially finds $2^{\frac{2k(N)-1}{k(N) \ell(N)}} \to_{N \to \infty} 1$ as long as $\ell(N)$ diverges. The result reported in Eq.~(\ref{eq:main_limit_results}) is obtained by noticing that
\begin{equation}
%\begin{array}{ll}
\begin{array}{l}
\lim_{N \to \infty}  \left[ 1 - 2^{-\frac{1}{k(N)}} \right]^{\frac{1}{\ell(N)}} = 
%&
\exp \left\{
\lim_{N \to \infty} \log  \left[ 1 - 2^{-\frac{1}{k(N)}} \right]^{\frac{1}{\ell(N)}} \right\} =
\\ 
%\exp \left\{ \lim_{N \to \infty} \frac{ \log \left[ 1 - 2^{-\frac{1}{k(N)}} %\right]}{\ell(N)} \right\} = 
%\\
% &
\exp \left\{ \lim_{N \to \infty} \frac{\log \log 2  - \log k(N)} {\ell(N)} \right\} =
%\\
% &
\exp \left\{ -  \lim_{N \to \infty} \frac{\log k(N)} {\ell(N)} \right\}
\; ,
\end{array}
\label{eq:limit_calculation}
\end{equation}
where the Taylor's series expansion $2^{-\frac{1}{k(N)}} = e^{-\frac{1}{k(N)} \log 2} \simeq 1 - \frac{1}{k(N)} \log 2 $ is used.
For any fixed $0 < E_{\textrm{target}} \leq 1$, replacing $2^{-1/k(N)}$ by $(1-E_{\textrm{target}}/2)^{1/k(N)}$ in the corresponding expression leads to the same asymptotic limits.

\newchange{
The other limit that is of interest in this paper concerns the redundancy generated by the heterogeneous protocols in the configuration model with power-law degree distribution, Eq.~(\ref{eq:degree_distribution}). For analytical estimates, we use the continuous approximation
\[
\mathcal{K}^{(0)}(\leq k)
=
\frac{k^{1-\gamma}-k_m^{1-\gamma}}
{k_M^{1-\gamma}-k_m^{1-\gamma}},
\qquad
\mathcal{Q}(\leq k)
=
\frac{k^{2-\gamma}-k_m^{2-\gamma}}
{k_M^{2-\gamma}-k_m^{2-\gamma}}.
\]

Equation~(\ref{eq:main_limit_results}) shows that, to sustain a non-vanishing level of entanglement as $\ell$ diverges, the redundancy $k$ must increase sufficiently rapidly with $\ell$. We therefore consider the probability that the redundancy of the channel exceeds a diverging scale $k(N)$.

For QPA, the redundancy is determined by the minimum degree of two randomly selected nodes. Since
\[
\lim_{N\to\infty}
\mathcal{K}^{(0)}_{\min}[\leq k(N)]=1
\]
for any $k(N)\to\infty$, the redundancy remains finite with probability approaching one. QPA is therefore not sustainable in the thermodynamic limit.

For h1QPA, the repeaters are selected among the neighbors of the end nodes. However,
\[
\lim_{N\to\infty}
\mathcal{Q}[\leq k(N)]=1
\]
for any $k(N)\to\infty$. It follows from the expression for
$\mathcal{K}^{(1)}_{\min}$ derived above that
\[
\lim_{N\to\infty}
\mathcal{K}^{(1)}_{\min}[\leq k(N)]=1.
\]
Thus, h1QPA is also not sustainable in the thermodynamic limit, although the approach to this limit can be extremely slow.

The situation is different for h2QPA. 
For $2<\gamma<3$, we have
$\mathcal{Q}(\leq N^\alpha)\sim
1-N^{\alpha(2-\gamma)}$ and
$S_2\sim\langle k^2\rangle\sim
N^{(3-\gamma)/2}$. Therefore,
\begin{equation}
\lim_{N\to\infty}
\left[\mathcal{Q}(\leq N^\alpha)\right]^{S_2(N)}
=0,
\qquad
0<\alpha<\frac{3-\gamma}{2}.
\label{eq:main_limit_results2}
\end{equation}
Consequently, the probability that the redundancy of the long-range channel grows at least as $N^\alpha$ does not vanish in the thermodynamic limit. Since $\ell\sim\log\log N$ in this regime, Eq.~(\ref{eq:main_limit_results}) then implies
$\log k/\ell\to\infty$ and therefore
$E_{\textrm{edge}}^*\to0$.

For $\gamma=3$, $S_2\sim\log N$. In this case,
\[
\lim_{N\to\infty}
\left[
\mathcal{Q}
\bigl(\leq(\log N)^\alpha\bigr)
\right]^{S_2(N)}
=0,
\qquad 0<\alpha<1.
\]
Since $\ell\sim\log\log N$, a redundancy
$k\sim(\log N)^\alpha$ gives
$\log k/\ell\to\alpha$, and Eq.~(\ref{eq:main_limit_results}) yields a finite threshold $E_{\textrm{edge}}^*<1$.

For $\gamma>3$, the second moment of the degree distribution, and hence $S_2$, remains finite as $N\to\infty$. Therefore,
\[
\lim_{N\to\infty}
\left[\mathcal{Q}(\leq k(N))\right]^{S_2(N)}
=1
\]
for every $k(N)\to\infty$, and h2QPA does not generate a sufficiently increasing redundancy. The same reasoning extends to $x>2$ by replacing $S_2$ with $S_x$, whose scaling is determined by successive powers of the excess-degree branching factor~\cite{newman2018networks}.

}

\begin{comment}
The other limit that is of interest in this paper is 
\begin{equation}
\lim_{N \to \infty} \left[ \mathcal{Q} (\leq N^{\alpha} )\right]^{S_2(N)} =  \left\{
\begin{array}{ll}
0 \textrm{ if } 2 < \gamma \leq 3
\\
1 \textrm{ if } \gamma > 3
\end{array}
\right. \; ,
    \label{eq:main_limit_results2}
\end{equation}
where $0 < \alpha < (3-\gamma)/2$.
This result follows from the fact that
$\mathcal{Q}(\leq N^{\alpha}) \sim 1 - N^{\alpha(2-\gamma)}$ and $S_2 \sim \langle k^2 \rangle \sim N^{(3-\gamma)/2}$ for $2 < \gamma < 3$ or $\langle k^2 \rangle \sim \log N$ for $\gamma =3$. Then, one can repeat a similar calculation as in Eq.~(\ref{eq:limit_calculation}). For $\gamma > 3$, $S_2(N)$ does not diverge with $N$, thus the limit is trivial. For $x > 2$, the very same limit applies, with the caveat that $S_x(N)$
is given by powers of the second moment of the degree distribution~\cite{newman2018networks}.

\end{comment}

\newchange{

\section{Numerical results for the configuration model}
\label{app:configuration_model}

\subsection{Network model and numerical implementation}

We consider networks generated according to the uncorrelated configuration model~\cite{molloy1995critical,catanzaro2005generation}. To create an instance of the model with $N$ nodes, we first extract $N$ random variables from the degree distribution
\begin{equation}
\mathcal{K}^{(0)}(k) \sim k^{-\gamma},
\qquad
k_m \leq k \leq k_M =
\min\left\{N^{1/2},N^{1/(\gamma-1)}\right\},
\label{eq:degree_distribution}
\end{equation}
with $\gamma>2$. We then create edges between nodes at random while preserving the prescribed degree sequence. Self-loops and multi-edges are suppressed. In all cases, we set $k_m=3$.

Numerical results are averaged over $10$ independent network instances for each combination of $N$ and $\gamma$, except for $N=10^8$, where we consider a single network instance.

\subsection{Degree of the repeaters}

The heterogeneous protocols exploit the largest-degree nodes found within distance $x$ from the end nodes. The distributions of these degrees provide a numerical illustration of the mechanism described in Appendix~B. Figure~\ref{fig:SMsynt_deg2} shows the distribution of the largest degree found within distance $x=1$ from a randomly selected node. For $2<\gamma\leq3$, the distribution shifts toward progressively larger values as $N$ increases, reflecting the increasing degree of the repeaters selected by h1QPA.

\begin{figure}[!htb]
    \centering
\includegraphics[width=0.45\textwidth]{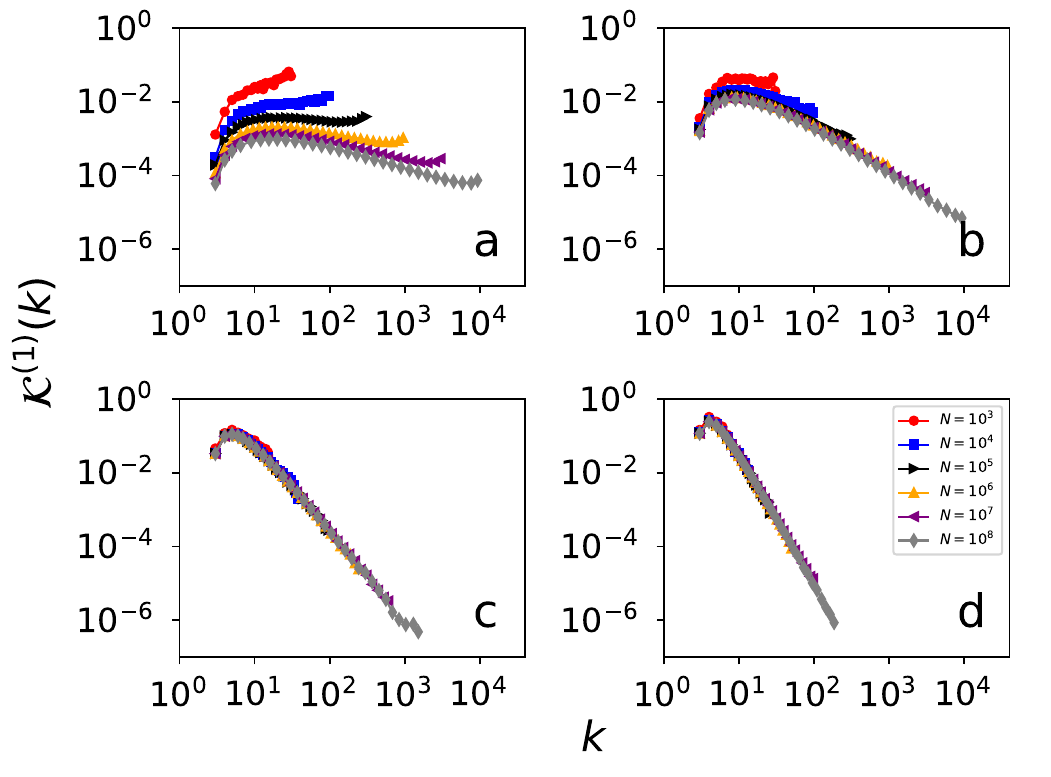}
    \caption{\textbf{Statistical properties of the configuration model.}
    (a) Distribution of the maximum degree of all nodes at most distance $x=1$
    from a randomly chosen node for networks of different sizes $N$.
    The degree exponent is $\gamma=2.1$.
    (b-d) Same as in (a), but for $\gamma=2.5$, $3.5$, and $4.5$, respectively.}
    \label{fig:SMsynt_deg2}
\end{figure}

The same analysis for $x=2$ is shown in Fig.~\ref{fig:SMsynt_deg4}. The increase with $N$ is substantially stronger in the regime $2<\gamma\leq3$, consistently with the analytical argument that the number of nodes available within the second neighborhood grows with network size and allows h2QPA to access increasingly high-degree repeaters. Due to the computational cost of evaluating the full second neighborhoods, for $N=10^8$ the distributions are estimated from a sample of $10^5$ randomly selected nodes.

\begin{figure}[!htb]
    \centering
\includegraphics[width=0.45\textwidth]{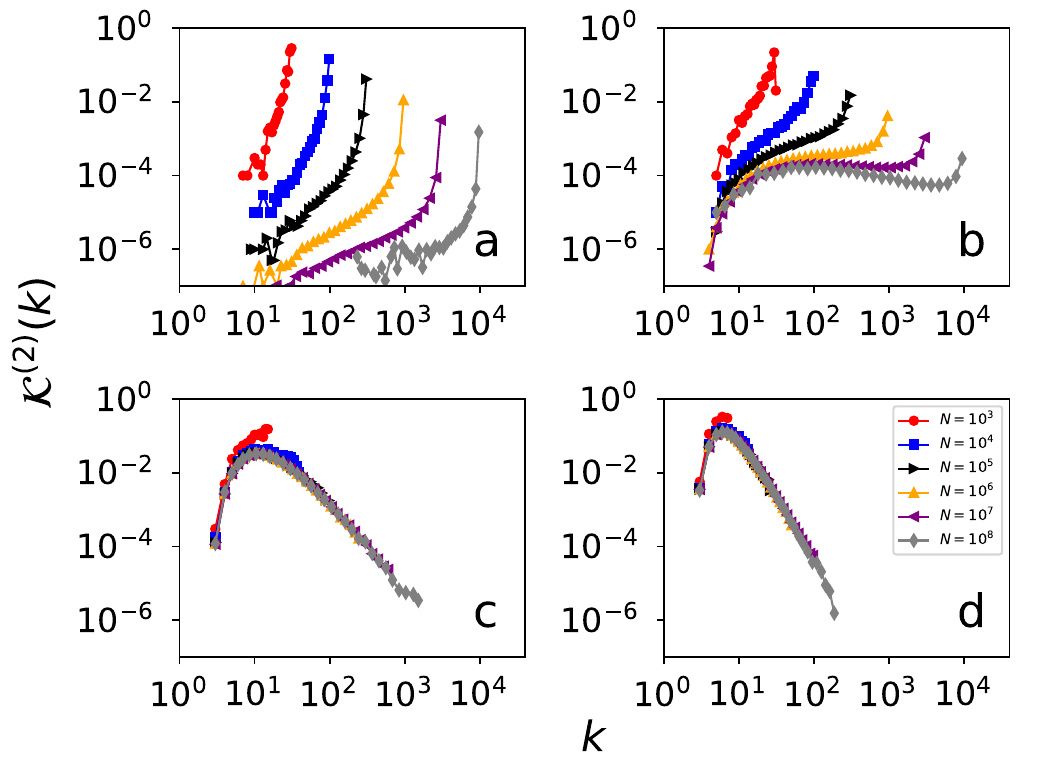}
    \caption{\textbf{Statistical properties of the configuration model.}
    (a) Distribution of the maximum degree of all nodes at most distance $x=2$
    from a randomly chosen node for networks of different sizes $N$.
    The degree exponent is $\gamma=2.1$.
    For $N=10^8$, the distribution is estimated from a sample of $10^5$
    randomly selected nodes.
    (b-d) Same as in (a), but for $\gamma=2.5$, $3.5$, and $4.5$, respectively.}
    \label{fig:SMsynt_deg4}
\end{figure}

\subsection{Performance of the communication protocols}

Figures~\ref{fig:SMsynth2.1}--\ref{fig:SMsynth4.5} show the average entanglement
$\langle E^{\textrm{prot}}\rangle$ as a function of the individual-edge
entanglement $E_{\textrm{edge}}$ for the four communication protocols and
different network sizes. These curves provide the numerical results from which
the AUC values displayed in Fig.~\ref{fig:2} are obtained.

Results for QPA and CPA are based on averages over $P=10000$ randomly selected
pairs of nodes. For $N=10^8$, we instead consider $P=1000$ pairs.
For h1QPA, we use at least $P=1000$ pairs, except for $N=10^8$, where
$P=100$; simulations for $N=10^8$ and $\gamma=2.1$ are not computationally
feasible. For h2QPA, we use at least $P=1000$ pairs for $N\leq10^5$ and
$P=100$ pairs for $N=10^6$ and $10^7$. 
Simulations for $N=10^8$ are not feasible. 
%only for $\gamma=4.5$, where we use $P=20$.

\begin{figure}[!htb]
    \centering
\includegraphics[width=0.45\textwidth]{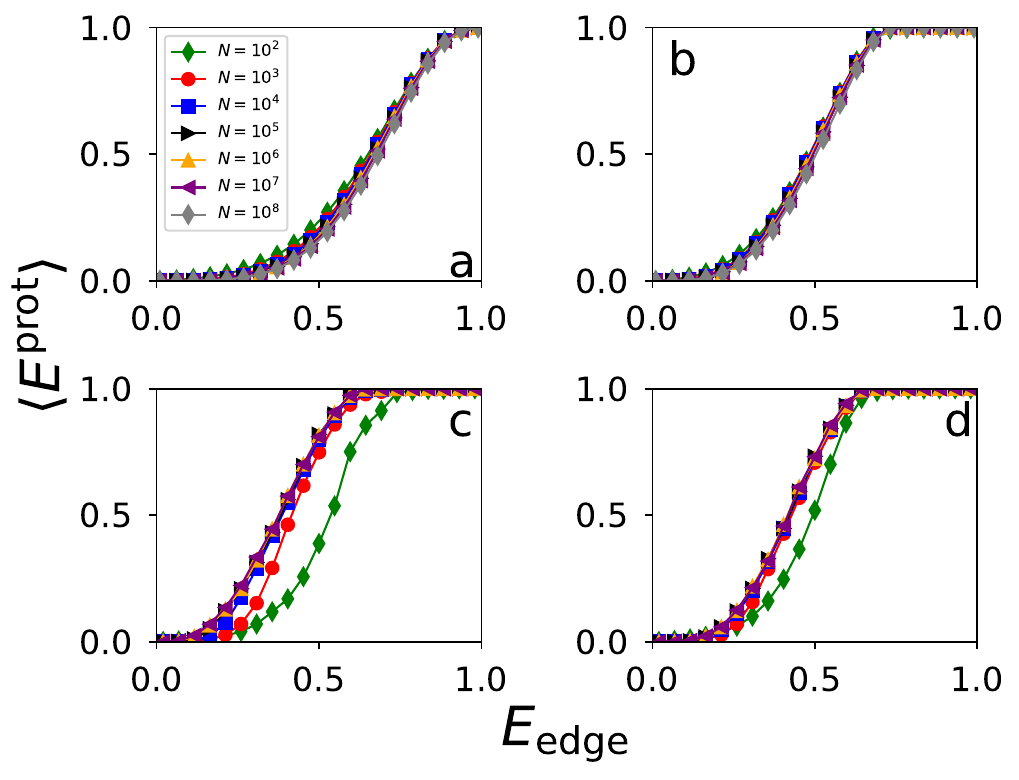}
    \caption{\textbf{Quantum communication on synthetic networks.}
    (a) Average entanglement $\langle E^{\textrm{CPA}}\rangle$ as a function
    of the entanglement of the individual edges $E_{\textrm{edge}}$ for
    networks generated according to the configuration model with degree
    exponent $\gamma=2.1$. Different symbols and colors correspond to
    different network sizes. (b-d) Same as in (a), but for QPA, h1QPA,
    and h2QPA, respectively.}
    \label{fig:SMsynth2.1}
\end{figure}

\begin{figure}[!htb]
    \centering
\includegraphics[width=0.45\textwidth]{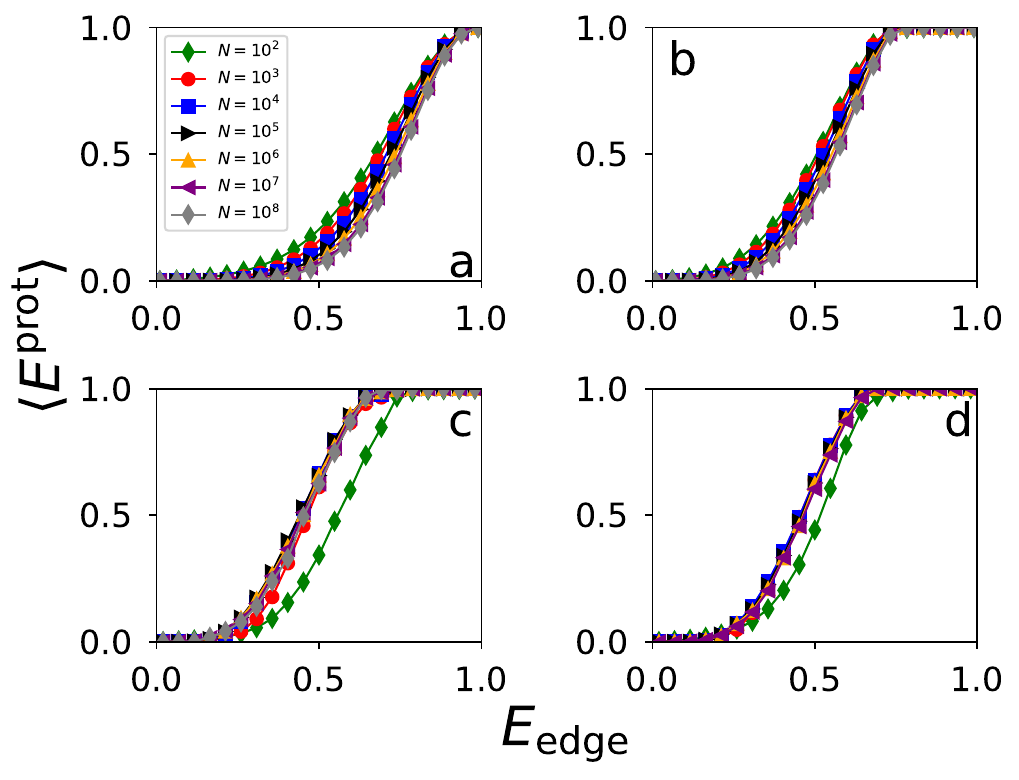}
    \caption{\textbf{Quantum communication on synthetic networks.}
    Same as in Fig.~\ref{fig:SMsynth2.1}, but for $\gamma=2.5$.}
    \label{fig:SMsynth2.5}
\end{figure}

\begin{figure}[!htb]
    \centering
\includegraphics[width=0.45\textwidth]{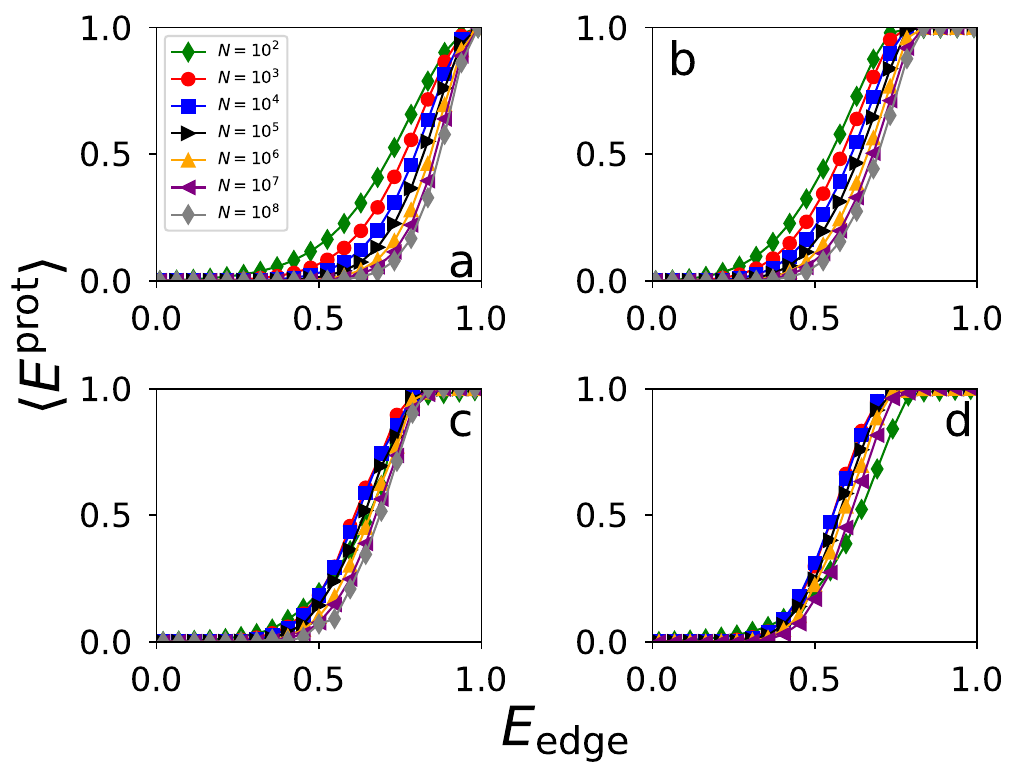}
    \caption{\textbf{Quantum communication on synthetic networks.}
    Same as in Fig.~\ref{fig:SMsynth2.1}, but for $\gamma=3.5$.}
    \label{fig:SMsynth3.5}
\end{figure}

\begin{figure}[!htb]
    \centering
\includegraphics[width=0.45\textwidth]{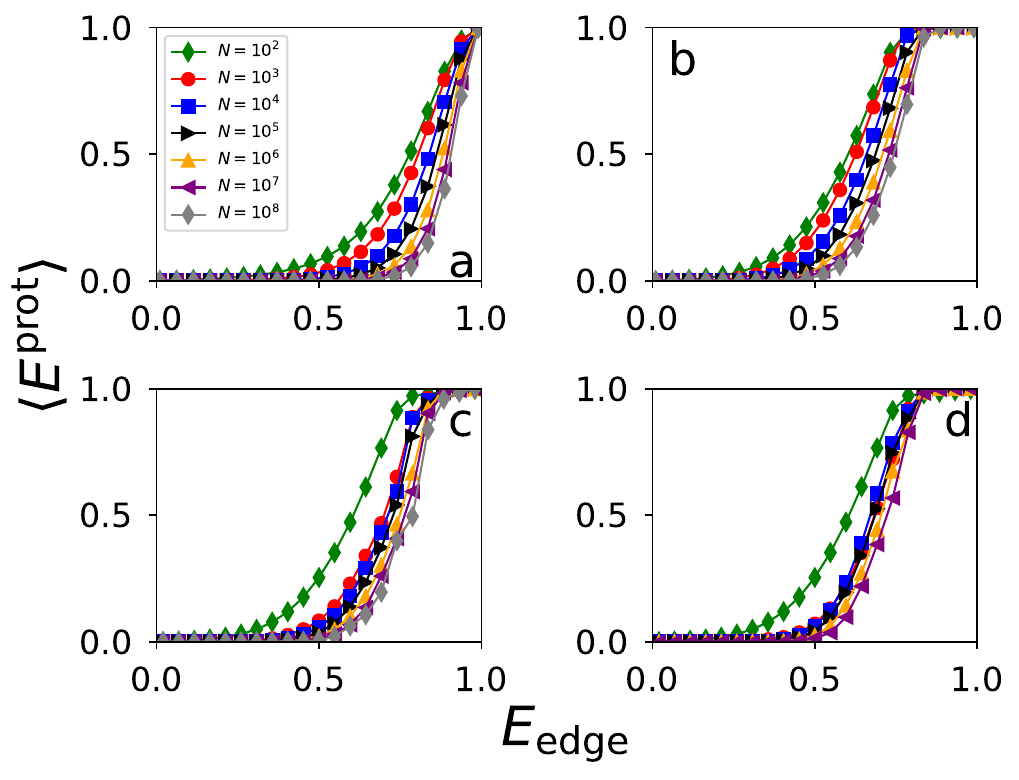}
    \caption{\textbf{Quantum communication on synthetic networks.}
    Same as in Fig.~\ref{fig:SMsynth2.1}, but for $\gamma=4.5$.}
    \label{fig:SMsynth4.5}
\end{figure}

\subsection{Computational cost}

We finally estimate the computational cost of the protocols by measuring the
time $T$ required to find all paths used to establish a communication channel
between two randomly selected nodes. Figure~\ref{fig:SMsynthtime} reports
results obtained in the nearly worst-case scenario $\lambda=0.99$, for which
the protocols exploit essentially all available paths. For larger levels of
initial edge entanglement, the computational time is generally smaller.

The numerical results display the approximate scaling
\begin{equation}
T\sim N^\eta.
\label{eq:time_scaling}
\end{equation}
For $2<\gamma\leq3$, the estimate of $\eta$ depends on the protocol, with the
heterogeneous variants displaying a stronger increase of computational cost
with $N$. For $\gamma>3$, the scaling becomes approximately independent of the
protocol.

\begin{figure}[!htb]
    \centering
\includegraphics[width=0.45\textwidth]{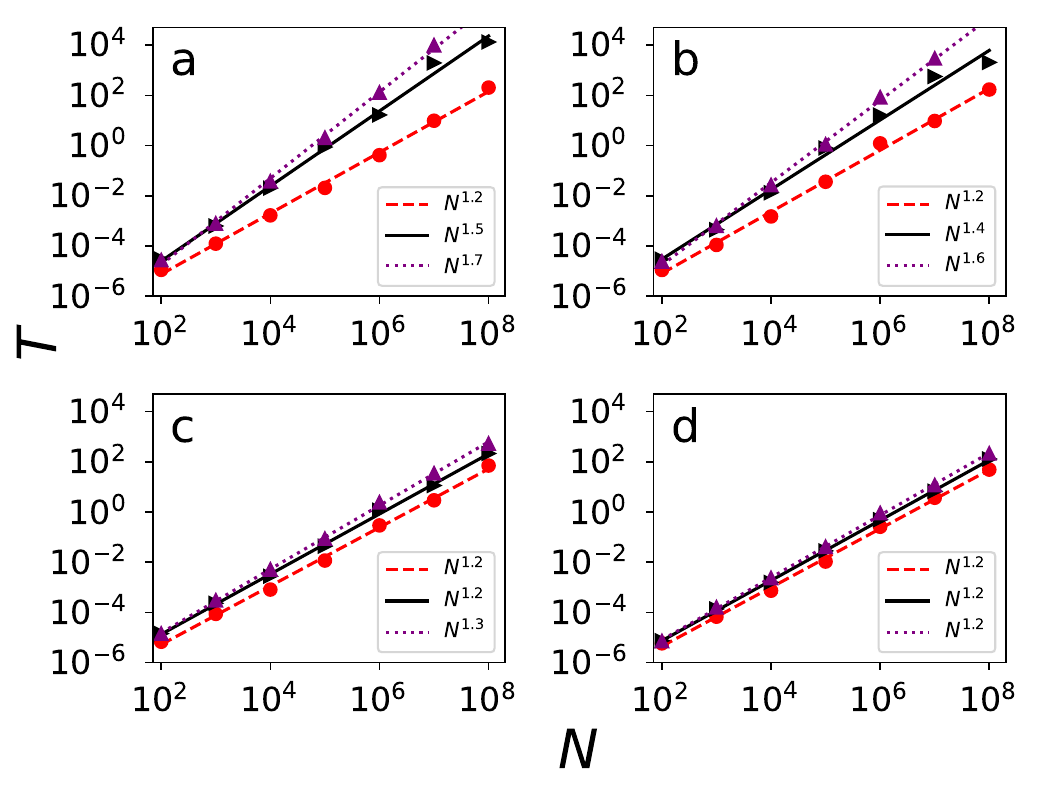}
    \caption{\textbf{Computational complexity of the communication protocols.}
    (a) Time $T$ required to find all paths between two randomly selected
    nodes $s$ and $t$ for QPA/CPA (red circles), h1QPA (black triangles),
    and h2QPA (purple triangles). Computations were performed with an
    \texttt{INTEL(R) XEON(R) GOLD 6548N} CPU. Time is measured in seconds
    and results are averaged over at least $10$ randomly selected pairs.
    Results are for the configuration model with $\gamma=2.1$.
    Lines represent power-law fits $T\sim N^\eta$; we obtain
    $\eta=1.2$ for QPA, $\eta=1.5$ for h1QPA, and $\eta=1.7$ for h2QPA.
    (b-d) Same as in (a), but for $\gamma=2.5$, $3.5$, and $4.5$,
    respectively. The corresponding estimates of $\eta$ are reported
    in the panels.}
    \label{fig:SMsynthtime}
\end{figure}

}

\newchange{

\section{Quantum communication on the $\mathbb{S}^1$ model}
\label{app:s1}

To test the robustness of our results against the presence of clustering, we consider networks generated according to the $\mathbb{S}^1$ model~\cite{serrano2008self}. We first assign to each node $i$ a degree $k_i$ by extracting a random variate from the distribution of Eq.~(\ref{eq:degree_distribution}), and an angular coordinate $\theta_i$ uniformly at random in the interval $[0,2\pi)$. Each pair of nodes $i$ and $j$ is then connected with probability
\[
p_{i,j}
=
\frac{1}{
1+
\left(
\frac{R\,\Delta\theta_{i,j}}
{\mu k_i k_j}
\right)^{-\beta}
},
\]
where $R=2\pi/N$, $\mu=(\beta-1)/(2\langle k\rangle)$,
$\langle k\rangle=N^{-1}\sum_i k_i$, and
$\Delta\theta_{i,j}
=
\min\{|\theta_i-\theta_j|,
2\pi-|\theta_i-\theta_j|\}$.
We consider $\beta=1.5$, which generates networks with average clustering coefficient approximately equal to $c=0.1$, with its precise value depending on $N$ and $\gamma$.

Numerical results are averaged over $5$ independent network instances for each combination of $N$ and $\gamma$, except for $N=10^7$, for which a single instance is considered. The quadratic time complexity of the generative algorithm prevents us from considering larger network sizes. The performance of the communication protocols is estimated using at least $P=1000$ randomly selected pairs of nodes.

Figures~\ref{fig:SMs1_2} and~\ref{fig:SMs1_4} show the distributions of the largest degree found within distance $x=1$ and $x=2$, respectively, from a randomly selected node. As for the configuration model, these quantities characterize the degree of the repeaters selected by h1QPA and h2QPA.

\begin{figure}[!htb]
\centering
\includegraphics[width=0.45\textwidth]{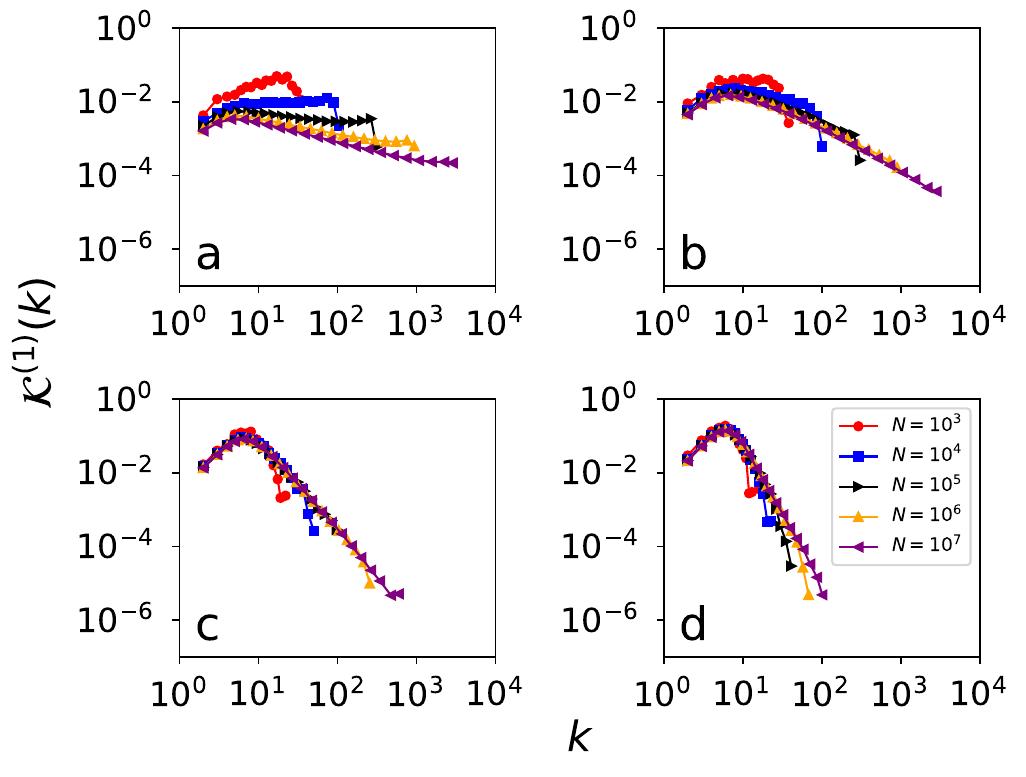}
\caption{\textbf{Statistical properties of the $\mathbb{S}^1$ model.}
(a) Distribution of the maximum degree among all nodes at most distance
$x=1$ from a randomly selected node for networks of different sizes $N$.
The degree exponent is $\gamma=2.1$.
(b-d) Same as in (a), but for $\gamma=2.5$, $3.5$, and $4.5$, respectively.}
\label{fig:SMs1_2}
\end{figure}

\begin{figure}[!htb]
\centering
\includegraphics[width=0.45\textwidth]{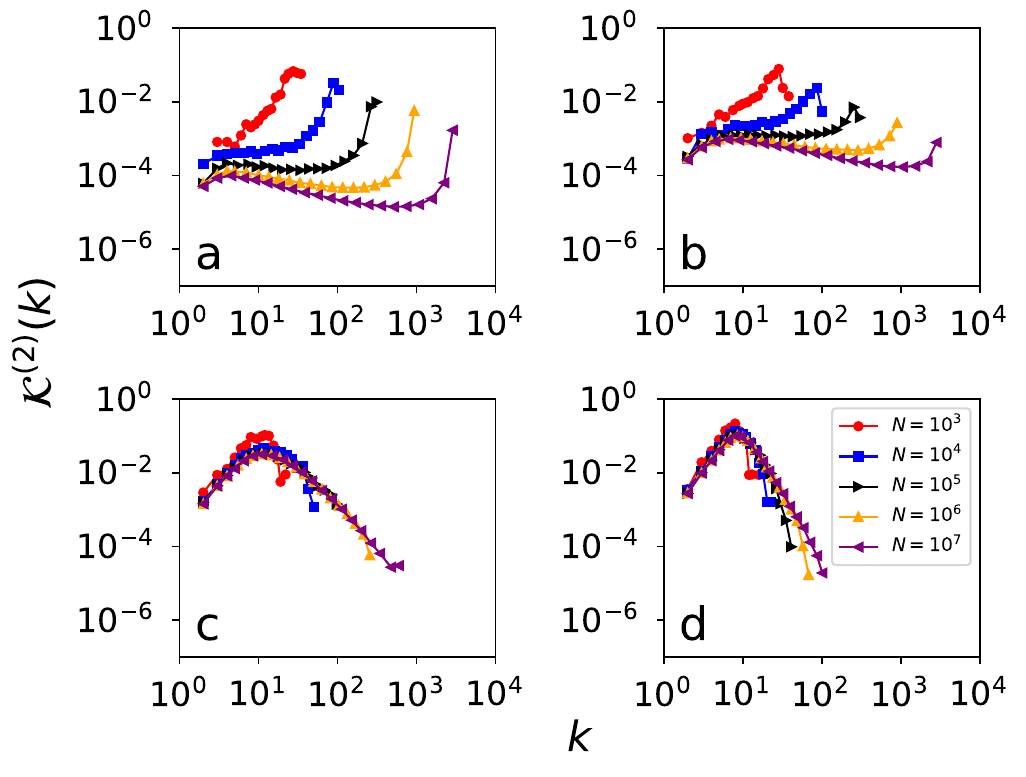}
\caption{\textbf{Statistical properties of the $\mathbb{S}^1$ model.}
(a) Distribution of the maximum degree among all nodes at most distance
$x=2$ from a randomly selected node for networks of different sizes $N$.
The degree exponent is $\gamma=2.1$.
(b-d) Same as in (a), but for $\gamma=2.5$, $3.5$, and $4.5$, respectively.}
\label{fig:SMs1_4}
\end{figure}

Figure~\ref{fig:SMs1} shows the AUC of the different communication protocols as a function of the network size. The results confirm that the qualitative behavior observed for the configuration model is not restricted to locally tree-like networks, but persists in networks with non-vanishing clustering.

\begin{figure}[!htb]
\centering
\includegraphics[width=0.45\textwidth]{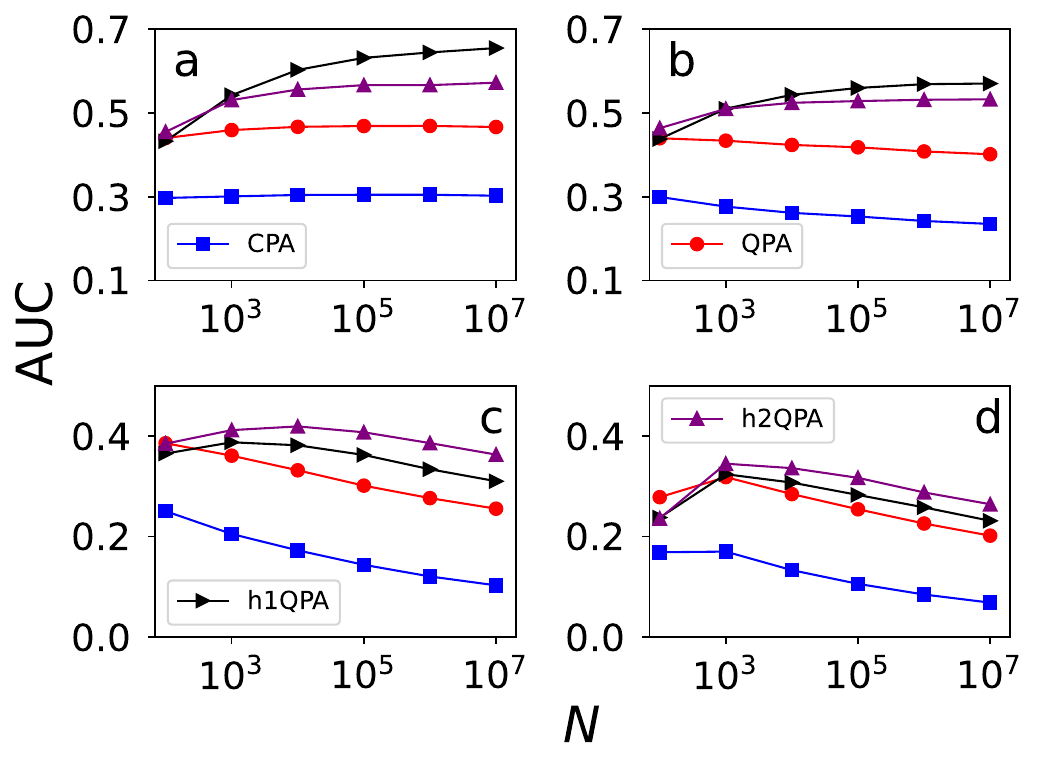}
\caption{\textbf{Quantum communication on the $\mathbb{S}^1$ model.}
(a) Area Under the Curve (AUC) for the various communication protocols
as a function of the network size $N$ for $\gamma=2.1$.
Results are averaged over multiple network instances and randomly selected
pairs of nodes $s$ and $t$.
(b-d) Same as in (a), but for $\gamma=2.5$, $3.5$, and $4.5$, respectively.}
\label{fig:SMs1}
\end{figure}

}

\clearpage

\newchange{

\section{Quantum communication on real networks}
\label{app:real_networks}

In Tables~\ref{tab:real1} and~\ref{tab:real2}, we provide a complete list of the real networks analyzed. Original network data have been preprocessed by removing eventual self-loops, multi edges, weights and directions to the edges; then, only the largest connected component of the resulting simple graph is considered in the analysis.
Results are obtained using $P=10000$ randomly selected pairs of nodes for QPA and CPA, and $P=1000$ for h1QPA and h2QPA.

\setlength{\abovecaptionskip}{10pt}

\begin{table*}[!htb]
    \centering
    \begin{tabular}{|l|r|r|r|r|r|r|r|c|r|}

 \hline
Network & $N$ & $M$ & $\langle k^2 \rangle/ \langle k \rangle^2$ & AUC$^{\textrm{CPA}}$
& AUC$^{\textrm{QPA}}$ & AUC$^{\textrm{h1QPA}}$
& AUC$^{\textrm{h2QPA}}$
& Refs.
& Data
\\
\hline \hline
        Social 3 & $32$ & $80$ & $1.19$ & $0.47$ & $0.61$ & $0.61$ & $0.62$ & \cite{milo2004superfamilies} & \cite{datasrc-uri-alon-lab-collection-of-complex-networks} \\ \hline

Karate club & $34$ & $78$ & $1.69$ & $0.43$ & $0.55$ & $0.61$ & $0.60$ & \cite{zachary1977information} & \cite{datasrc-mark-e-j-newman-network-data} \\ \hline

Protein 2 & $53$ & $123$ & $1.16$ & $0.33$ & $0.48$ & $0.47$ & $0.49$ & \cite{milo2004superfamilies} & \cite{datasrc-uri-alon-lab-collection-of-complex-networks} \\ \hline

Dolphins & $62$ & $159$ & $1.33$ & $0.36$ & $0.49$ & $0.51$ & $0.52$ & \cite{lusseau2003bottlenose} & \cite{datasrc-mark-e-j-newman-network-data} \\ \hline

Social 1 & $67$ & $142$ & $1.24$ & $0.34$ & $0.48$ & $0.48$ & $0.50$ & \cite{milo2004superfamilies} & \cite{datasrc-uri-alon-lab-collection-of-complex-networks} \\ \hline

Les Miserables & $77$ & $254$ & $1.83$ & $0.39$ & $0.50$ & $0.59$ & $0.54$ & \cite{knuth1993stanford} & \cite{datasrc-mark-e-j-newman-network-data} \\ \hline

Protein 1 & $95$ & $213$ & $1.10$ & $0.22$ & $0.35$ & $0.33$ & $0.33$ & \cite{milo2004superfamilies} & \cite{datasrc-uri-alon-lab-collection-of-complex-networks} \\ \hline

E. Coli, transcription & $97$ & $212$ & $1.24$ & $0.25$ & $0.40$ & $0.39$ & $0.42$ & \cite{mangan2003structure} & \cite{datasrc-uri-alon-lab-collection-of-complex-networks} \\ \hline

Political books & $105$ & $441$ & $1.42$ & $0.44$ & $0.60$ & $0.66$ & $0.64$ & \cite{adamic2005political} & \cite{datasrc-mark-e-j-newman-network-data} \\ \hline

David Copperfield & $112$ & $425$ & $1.81$ & $0.45$ & $0.57$ & $0.65$ & $0.60$ & \cite{newman2006finding} & \cite{datasrc-mark-e-j-newman-network-data} \\ \hline

College football & $115$ & $613$ & $1.01$ & $0.55$ & $0.70$ & $0.62$ & $0.64$ & \cite{girvan2002community} & \cite{datasrc-mark-e-j-newman-network-data} \\ \hline

S 208 & $122$ & $189$ & $1.22$ & $0.24$ & $0.38$ & $0.34$ & $0.38$ & \cite{milo2004superfamilies} & \cite{datasrc-uri-alon-lab-collection-of-complex-networks} \\ \hline

High school, 2011 & $126$ & $1709$ & $1.21$ & $0.70$ & $0.80$ & $0.83$ & $0.81$ & \cite{fournet2014contact} & \cite{datasrc-sociopatterns-collaboration-high-school-dynamic-contact-networks} \\ \hline

Bay Dry & $128$ & $2106$ & $1.23$ & $0.75$ & $0.84$ & $0.86$ & $0.84$ & \cite{ulanowicz1998network, konect} & \cite{datasrc-konect-bay-dry-network-dataset} \\ \hline

Bay Wet & $128$ & $2075$ & $1.24$ & $0.75$ & $0.84$ & $0.86$ & $0.84$ & \cite{konect} & \cite{datasrc-konect-bay-wet-network-dataset} \\ \hline

Radoslaw Email & $167$ & $3250$ & $1.66$ & $0.64$ & $0.71$ & $0.76$ & $0.72$ & \cite{radoslaw, konect} & \cite{datasrc-konect-radoslaw-email-network-dataset} \\ \hline

High school, 2012 & $180$ & $2220$ & $1.20$ & $0.66$ & $0.78$ & $0.81$ & $0.79$ & \cite{fournet2014contact} & \cite{datasrc-sociopatterns-collaboration-high-school-dynamic-contact-networks} \\ \hline
Little Rock Lake & $183$ & $2434$ & $1.61$ & $0.63$ & $0.74$ & $0.78$ & $0.77$ & \cite{martinez1991artifacts, konect} & \cite{datasrc-konect-little-rock-lake-network-dataset} \\ \hline
Jazz & $198$ & $2742$ & $1.40$ & $0.62$ & $0.74$ & $0.81$ & $0.76$ & \cite{gleiser2003community} & \cite{datasrc-alexandre-arenas-network-datasets} \\ \hline
S 420 & $252$ & $399$ & $1.23$ & $0.21$ & $0.36$ & $0.34$ & $0.37$ & \cite{milo2004superfamilies} & \cite{datasrc-uri-alon-lab-collection-of-complex-networks} \\ \hline
C. Elegans, neural & $297$ & $2148$ & $1.80$ & $0.53$ & $0.66$ & $0.74$ & $0.70$ & \cite{watts1998collective} & \cite{datasrc-mark-e-j-newman-network-data} \\ \hline
Network Science & $379$ & $914$ & $1.66$ & $0.20$ & $0.35$ & $0.41$ & $0.39$ & \cite{newman2006finding} & \cite{datasrc-mark-e-j-newman-network-data} \\ \hline

   Dublin & $410$ & $2765$ & $1.39$ & $0.42$ & $0.58$ & $0.66$ & $0.66$ & \cite{isella2011s, konect} & \cite{datasrc-konect-dublin-network-dataset} \\ \hline

US Air Trasportation & $500$ & $2980$ & $4.51$ & $0.36$ & $0.48$ & $0.59$ & $0.53$ & \cite{colizza2007reaction} & \cite{datasrc-tore-opsahl-us-top-500-airport-network-dataset} \\ \hline

S 838 & $512$ & $819$ & $1.26$ & $0.18$ & $0.33$ & $0.33$ & $0.35$ & \cite{milo2004superfamilies} & \cite{datasrc-uri-alon-lab-collection-of-complex-networks} \\ \hline

Yeast, transcription & $662$ & $1062$ & $4.21$ & $0.20$ & $0.31$ & $0.39$ & $0.39$ & \cite{milo2002network} & \cite{datasrc-uri-alon-lab-collection-of-complex-networks} \\ \hline

URV email & $1133$ & $5451$ & $1.94$ & $0.35$ & $0.50$ & $0.60$ & $0.56$ & \cite{guimera2003self} & \cite{datasrc-alexandre-arenas-network-datasets} \\ \hline

Political blogs & $1222$ & $16714$ & $2.97$ & $0.47$ & $0.59$ & $0.70$ & $0.64$ & \cite{adamic2005political} & \cite{datasrc-mark-e-j-newman-network-data} \\ \hline

Air traffic & $1226$ & $2408$ & $1.87$ & $0.21$ & $0.36$ & $0.41$ & $0.43$ & \cite{konect} & \cite{datasrc-konect-air-traffic-network-dataset} \\ \hline

Yeast, protein & $1458$ & $1948$ & $2.67$ & $0.15$ & $0.26$ & $0.29$ & $0.31$ & \cite{jeong2001lethality} & \cite{datasrc-network-data-repository-yeast-protein-network-dataset} \\ \hline

Petster, hamster & $1788$ & $12476$ & $3.26$ & $0.38$ & $0.51$ & $0.63$ & $0.58$ & \cite{konect} & \cite{datasrc-konect-petster-hamster-network-dataset} \\ \hline

UC Irvine & $1893$ & $13835$ & $3.81$ & $0.38$ & $0.50$ & $0.61$ & $0.55$ & \cite{opsahl2009clustering, konect} & \cite{datasrc-konect-uc-irvine-network-dataset} \\ \hline

Yeast, protein & $2224$ & $6609$ & $2.80$ & $0.25$ & $0.38$ & $0.49$ & $0.45$ & \cite{bu2003topological} & \cite{datasrc-pajek-datasets-yeast-protein-protein-interaction-network} \\ \hline

Japanese & $2698$ & $7995$ & $18.33$ & $0.31$ & $0.42$ & $0.52$ & $0.47$ & \cite{milo2004superfamilies} & \cite{datasrc-uri-alon-lab-collection-of-complex-networks} \\ \hline

Open flights & $2905$ & $15645$ & $5.18$ & $0.28$ & $0.41$ & $0.55$ & $0.48$ & \cite{opsahl2010node, konect} & \cite{datasrc-konect-open-flights-network-dataset} \\ \hline

GR-QC, 1993-2003 & $4158$ & $13422$ & $2.79$ & $0.20$ & $0.34$ & $0.44$ & $0.43$ & \cite{leskovec2007graph} & \cite{datasrc-stanford-network-analysis-project-gr-qc-1993-2003-network-dataset} \\ \hline

Tennis & $4338$ & $81865$ & $4.21$ & $0.34$ & $0.46$ & $0.59$ & $0.52$ & \cite{radicchi2011best} & - \\ \hline

US Power grid & $4941$ & $6594$ & $1.45$ & $0.07$ & $0.18$ & $0.18$ & $0.19$ & \cite{watts1998collective} & \cite{datasrc-mark-e-j-newman-network-data} \\ \hline

HT09 & $5352$ & $18481$ & $28.96$ & $0.29$ & $0.41$ & $0.54$ & $0.52$ & \cite{isella2011s} & \cite{datasrc-sociopatterns-collaboration-hypertext-2009-dynamic-contact-network} \\ \hline

Hep-Th, 1995-1999 & $5835$ & $13815$ & $1.93$ & $0.17$ & $0.32$ & $0.40$ & $0.41$ & \cite{newman2001structure} & \cite{datasrc-mark-e-j-newman-network-data} \\ \hline

Reactome & $5973$ & $145778$ & $2.94$ & $0.39$ & $0.54$ & $0.66$ & $0.63$ & \cite{joshi2005reactome, konect} & \cite{datasrc-konect-reactome-network-dataset} \\ \hline

Jung & $6120$ & $50290$ & $60.35$ & $0.55$ & $0.68$ & $0.74$ & $0.69$ & \cite{vsubelj2012software, konect} & \cite{datasrc-konect-jung-network-dataset} \\ \hline

Gnutella, Aug. 8, 2002 & $6299$ & $20776$ & $2.68$ & $0.25$ & $0.39$ & $0.48$ & $0.46$ & \cite{ripeanu2002mapping, leskovec2007graph} & \cite{datasrc-stanford-network-analysis-project-gnutella-aug-8-2002-network-dataset} \\ \hline

JDK & $6434$ & $53658$ & $58.92$ & $0.55$ & $0.68$ & $0.75$ & $0.69$ & \cite{konect} & \cite{datasrc-konect-jdk-network-dataset} \\ \hline

AS Oregon$^{*}$ & $6474$ & $12572$ & $42.43$ & $0.26$ & $0.37$ & $0.48$ & $0.43$ & \cite{leskovec2005graphs} & \cite{datasrc-stanford-network-analysis-project-as-oregon-network-dataset} \\ \hline

English & $7377$ & $44205$ & $26.76$ & $0.38$ & $0.50$ & $0.61$ & $0.54$ & \cite{milo2004superfamilies} & \cite{datasrc-uri-alon-lab-collection-of-complex-networks} \\ \hline

WHOIS$^{*}$ & $7476$ & $56943$ & $8.91$ & $0.33$ & $0.47$ & $0.62$ & $0.54$ & \cite{nr, mahadevan2006internet} & \cite{datasrc-network-data-repository-whois-network-dataset} \\ \hline

Gnutella, Aug. 9, 2002 & $8104$ & $26008$ & $2.62$ & $0.24$ & $0.38$ & $0.48$ & $0.45$ & \cite{ripeanu2002mapping, leskovec2007graph} & \cite{datasrc-stanford-network-analysis-project-gnutella-aug-9-2002-network-dataset} \\ \hline

French & $8308$ & $23832$ & $38.00$ & $0.30$ & $0.42$ & $0.53$ & $0.47$ & \cite{milo2004superfamilies} & \cite{datasrc-uri-alon-lab-collection-of-complex-networks} \\ \hline

Hep-Th, 1993-2003 & $8638$ & $24806$ & $2.26$ & $0.21$ & $0.35$ & $0.46$ & $0.45$ & \cite{leskovec2007graph} & \cite{datasrc-stanford-network-analysis-project-hep-th-1993-2003-network-dataset} \\ \hline

nutella, Aug. 6, 2002 & $8717$ & $31525$ & $1.99$ & $0.27$ & $0.41$ & $0.50$ & $0.49$ & \cite{ripeanu2002mapping, leskovec2007graph} & \cite{datasrc-stanford-network-analysis-project-nutella-aug-6-2002-network-dataset} \\ \hline

Gnutella, Aug. 5, 2002 & $8842$ & $31837$ & $2.05$ & $0.27$ & $0.41$ & $0.50$ & $0.49$ & \cite{ripeanu2002mapping, leskovec2007graph} & \cite{datasrc-stanford-network-analysis-project-gnutella-aug-5-2002-network-dataset} \\ \hline

PGP & $10680$ & $24316$ & $4.15$ & $0.14$ & $0.25$ & $0.31$ & $0.32$ & \cite{boguna2004models} & \cite{datasrc-alexandre-arenas-network-datasets} \\ \hline

Gnutella, August 4 2002 & $10876$ & $39994$ & $1.90$ & $0.27$ & $0.41$ & $0.50$ & $0.48$ & \cite{ripeanu2002mapping, leskovec2007graph} & \cite{datasrc-stanford-network-analysis-project-gnutella-august-4-2002-network-dataset} \\ \hline

Hep-Ph, 1993-2003 & $11204$ & $117619$ & $6.23$ & $0.29$ & $0.44$ & $0.59$ & $0.53$ & \cite{leskovec2007graph} & \cite{datasrc-stanford-network-analysis-project-hep-ph-1993-2003-network-dataset} \\ \hline

Spanish & $11558$ & $43050$ & $61.43$ & $0.32$ & $0.43$ & $0.54$ & $0.47$ & \cite{milo2004superfamilies} & \cite{datasrc-uri-alon-lab-collection-of-complex-networks} \\ \hline

DBLP, citations & $12495$ & $49563$ & $5.52$ & $0.24$ & $0.37$ & $0.51$ & $0.45$ & \cite{ley2002dblp, konect} & \cite{datasrc-konect-dblp-citations-network-dataset} \\ \hline

Spanish & $12643$ & $55019$ & $92.80$ & $0.38$ & $0.51$ & $0.61$ & $0.58$ & \cite{konect} & \cite{datasrc-konect-spanish-network-dataset} \\ \hline

Cond-Mat, 1995-1999 & $13861$ & $44619$ & $2.10$ & $0.20$ & $0.36$ & $0.46$ & $0.47$ & \cite{newman2001structure} & \cite{datasrc-mark-e-j-newman-network-data} \\ \hline

Astrophysics & $14845$ & $119652$ & $2.82$ & $0.31$ & $0.47$ & $0.62$ & $0.57$ & \cite{newman2001structure} & \cite{datasrc-mark-e-j-newman-network-data} \\ \hline

Google & $15763$ & $148585$ & $47.83$ & $0.44$ & $0.56$ & $0.69$ & $0.60$ & \cite{palla2007directed} & \cite{datasrc-konect-google-network-dataset} \\ \hline

AstroPhys, 1993-2003 & $17903$ & $196972$ & $2.99$ & $0.36$ & $0.52$ & $0.67$ & $0.61$ & \cite{leskovec2007graph} & \cite{datasrc-stanford-network-analysis-project-astrophys-1993-2003-network-dataset} \\ \hline
    
    \end{tabular}
    \caption{\textbf{Real networks.} From left to right, for each network we report: name, number of nodes, number of edges, ratio of the second moment of the degree distribution and the square of the first moment, AUC metrics for the protocols CPA, QPA, h1QPA and h2QPA, paper(s) where the dataset was first studied, and numbered reference to the data source. We consider only the largest connected component of each network. Snapshot of the Internet are highlighted by an asterisk close to the network name.}
    \label{tab:real1}
\end{table*}

\begin{table*}[!htb]
    \centering
    \begin{tabular}{|l|r|r|r|r|r|r|r|c|r|}
 \hline
Network & $N$ & $M$ & $\langle k^2 \rangle/ \langle k \rangle^2$ & AUC$^{\textrm{CPA}}$
& AUC$^{\textrm{QPA}}$ & AUC$^{\textrm{h1QPA}}$
& AUC$^{\textrm{h2QPA}}$
& Refs.
& Data
\\
\hline \hline

Cond-Mat, 1993-2003 & $21363$ & $91286$ & $2.63$ & $0.26$ & $0.42$ & $0.56$ & $0.53$ & \cite{leskovec2007graph} & \cite{datasrc-stanford-network-analysis-project-cond-mat-1993-2003-network-dataset} \\ \hline

Gnutella, Aug. 25, 2002 & $22663$ & $54693$ & $2.23$ & $0.20$ & $0.32$ & $0.43$ & $0.40$ & \cite{ripeanu2002mapping, leskovec2007graph} & \cite{datasrc-stanford-network-analysis-project-gnutella-aug-25-2002-network-dataset} \\ \hline

Internet & $22963$ & $48436$ & $61.98$ & $0.25$ & $0.37$ & $0.49$ & $0.42$ & - & \cite{datasrc-mark-e-j-newman-network-data} \\ \hline

Thesaurus & $23132$ & $297094$ & $4.02$ & $0.38$ & $0.52$ & $0.64$ & $0.58$ & \cite{kiss1973associative, konect} & \cite{datasrc-konect-thesaurus-network-dataset} \\ \hline

Cora & $23166$ & $89157$ & $3.08$ & $0.23$ & $0.39$ & $0.51$ & $0.51$ & \cite{vsubelj2013model, konect} & \cite{datasrc-konect-cora-network-dataset} \\ \hline

Linux, mailing list & $24567$ & $158164$ & $26.48$ & $0.25$ & $0.37$ & $0.45$ & $0.43$ & \cite{konect} & \cite{datasrc-konect-linux-mailing-list-network-dataset} \\ \hline

AS Caida$^{*}$ & $26475$ & $53381$ & $69.50$ & $0.25$ & $0.36$ & $0.48$ & $0.41$ & \cite{leskovec2005graphs} & \cite{datasrc-stanford-network-analysis-project-as-caida-network-dataset} \\ \hline

Gnutella, Aug. 24, 2002 & $26498$ & $65359$ & $2.44$ & $0.20$ & $0.32$ & $0.44$ & $0.41$ & \cite{ripeanu2002mapping, leskovec2007graph} & \cite{datasrc-stanford-network-analysis-project-gnutella-aug-24-2002-network-dataset} \\ \hline

Hep-Th, citations & $27400$ & $352021$ & $4.14$ & $0.39$ & $0.55$ & $0.67$ & $0.64$ & \cite{leskovec2007graph, konect} & \cite{datasrc-konect-hep-th-citations-network-dataset} \\ \hline

Cond-Mat, 1995-2003 & $27519$ & $116181$ & $2.64$ & $0.24$ & $0.40$ & $0.54$ & $0.52$ & \cite{newman2001structure} & \cite{datasrc-mark-e-j-newman-network-data} \\ \hline

Digg & $29652$ & $84781$ & $4.91$ & $0.22$ & $0.35$ & $0.47$ & $0.42$ & \cite{de2009social, konect} & \cite{datasrc-konect-digg-network-dataset} \\ \hline

Linux, soft. & $30817$ & $213208$ & $61.62$ & $0.39$ & $0.53$ & $0.64$ & $0.60$ & \cite{konect} & \cite{datasrc-konect-linux-soft-network-dataset} \\ \hline

Enron & $33696$ & $180811$ & $13.27$ & $0.27$ & $0.40$ & $0.55$ & $0.48$ & \cite{leskovec2009community} & \cite{datasrc-stanford-network-analysis-project-enron-network-dataset} \\ \hline

Hep-Ph, citations & $34401$ & $420784$ & $2.60$ & $0.39$ & $0.56$ & $0.71$ & $0.67$ & \cite{leskovec2007graph, konect} & \cite{datasrc-konect-hep-ph-citations-network-dataset} \\ \hline

Cond-Mat, 1995-2005 & $36458$ & $171735$ & $2.96$ & $0.25$ & $0.42$ & $0.57$ & $0.53$ & \cite{newman2001structure} & \cite{datasrc-mark-e-j-newman-network-data} \\ \hline

Gnutella, Aug. 30, 2002 & $36646$ & $88303$ & $2.38$ & $0.19$ & $0.31$ & $0.43$ & $0.39$ & \cite{ripeanu2002mapping, leskovec2007graph} & \cite{datasrc-stanford-network-analysis-project-gnutella-aug-30-2002-network-dataset} \\ \hline

AS Internet$^{*}$ & $40164$ & $85123$ & $79.53$ & $0.24$ & $0.36$ & $0.48$ & $0.41$ & \cite{nr, rossi2013topology} & \cite{datasrc-network-data-repository-as-internet-network-dataset} \\ \hline

Slashdot & $51083$ & $116573$ & $17.87$ & $0.20$ & $0.31$ & $0.42$ & $0.37$ & \cite{gomez2008statistical, konect} & \cite{datasrc-konect-slashdot-network-dataset} \\ \hline

Gnutella, Aug. 31, 2002 & $62561$ & $147878$ & $2.45$ & $0.18$ & $0.30$ & $0.42$ & $0.39$ & \cite{ripeanu2002mapping, leskovec2007graph} & \cite{datasrc-stanford-network-analysis-project-gnutella-aug-31-2002-network-dataset} \\ \hline

Gnutella & $62561$ & $147878$ & $2.45$ & $0.18$ & $0.30$ & $0.42$ & $0.39$ & \cite{nr, matei2002mapping} & \cite{datasrc-network-data-repository-gnutella-network-dataset} \\ \hline

Facebook & $63392$ & $816886$ & $3.42$ & $0.34$ & $0.49$ & $0.63$ & $0.58$ & \cite{viswanath2009evolution} & \cite{datasrc-max-planck-institute-for-software-systems-wosn-2009-facebook-dataset} \\ \hline

Epinions & $75877$ & $405739$ & $17.19$ & $0.23$ & $0.33$ & $0.46$ & $0.40$ & \cite{richardson2003trust, konect} & \cite{datasrc-konect-epinions-network-dataset} \\ \hline

Slashdot zoo & $79116$ & $467731$ & $12.37$ & $0.26$ & $0.38$ & $0.51$ & $0.44$ & \cite{kunegis2009slashdot, konect} & \cite{datasrc-konect-slashdot-zoo-network-dataset} \\ \hline

Flickr & $105722$ & $2316668$ & $7.97$ & $0.37$ & $0.55$ & $0.66$ & $0.64$ & \cite{McAuley2012, konect} & \cite{datasrc-konect-flickr-network-dataset} \\ \hline

Wikipedia, edits & $113123$ & $2025910$ & $19.25$ & $0.35$ & $0.48$ & $0.62$ & $0.54$ & \cite{brandes2010structural, konect} & \cite{datasrc-konect-wikipedia-edits-network-dataset} \\ \hline

Router-level Caida$^{*}$ & $190914$ & $607610$ & $5.96$ & $0.17$ & $0.31$ & $0.42$ & $0.43$ & \cite{nr, rossi2013topology, spring2002measuring} & \cite{datasrc-network-data-repository-router-level-caida-network-dataset} \\ \hline

Gowalla & $196591$ & $950327$ & $31.71$ & $0.25$ & $0.38$ & $0.51$ & $0.47$ & \cite{cho2011friendship, konect} & \cite{datasrc-konect-gowalla-network-dataset} \\ \hline

EU email & $224832$ & $339925$ & $187.73$ & $0.20$ & $0.30$ & $0.41$ & $0.33$ & \cite{leskovec2007graph, konect} & \cite{datasrc-konect-eu-email-network-dataset} \\ \hline

Web Stanford & $255265$ & $1941926$ & $133.47$ & $0.22$ & $0.37$ & $0.47$ & $0.47$ & \cite{leskovec2009community} & \cite{datasrc-stanford-network-analysis-project-web-stanford-network-dataset} \\ \hline

Amazon, Mar. 2, 2003 & $262111$ & $899792$ & $1.62$ & $0.18$ & $0.37$ & $0.43$ & $0.47$ & \cite{leskovec2007dynamics} & \cite{datasrc-stanford-network-analysis-project-amazon-mar-2-2003-network-dataset} \\ \hline

DBLP, collaborations & $317080$ & $1049866$ & $3.28$ & $0.19$ & $0.34$ & $0.48$ & $0.46$ & \cite{ley2002dblp, konect} & \cite{datasrc-konect-dblp-collaborations-network-dataset} \\ \hline

Web Notre Dame & $325729$ & $1090108$ & $41.93$ & $0.15$ & $0.26$ & $0.34$ & $0.33$ & \cite{albert1999internet} & \cite{datasrc-stanford-network-analysis-project-web-notre-dame-network-dataset} \\ \hline

MathSciNet & $332689$ & $820644$ & $3.33$ & $0.16$ & $0.29$ & $0.41$ & $0.41$ & \cite{palla2008fundamental} & \cite{datasrc-network-data-repository-mathscinet-network-dataset} \\ \hline

CiteSeer & $365154$ & $1721981$ & $5.14$ & $0.22$ & $0.38$ & $0.52$ & $0.51$ & \cite{bollacker1998citeseer, konect} & \cite{datasrc-konect-citeseer-network-dataset} \\ \hline

Zhishi & $372840$ & $2318025$ & $2244.54$ & $0.27$ & $0.42$ & $0.51$ & $0.53$ & \cite{niu2011zhishi, konect} & \cite{datasrc-konect-zhishi-network-dataset} \\ \hline

Amazon, Mar. 12, 2003 & $400727$ & $2349869$ & $2.59$ & $0.26$ & $0.45$ & $0.56$ & $0.58$ & \cite{leskovec2007dynamics} & \cite{datasrc-stanford-network-analysis-project-amazon-mar-12-2003-network-dataset} \\ \hline

Amazon, Jun. 6, 2003 & $403364$ & $2443311$ & $2.52$ & $0.27$ & $0.46$ & $0.57$ & $0.59$ & \cite{leskovec2007dynamics} & \cite{datasrc-stanford-network-analysis-project-amazon-jun-6-2003-network-dataset} \\ \hline

Amazon, May 5, 2003 & $410236$ & $2439437$ & $2.60$ & $0.26$ & $0.45$ & $0.56$ & $0.59$ & \cite{leskovec2007dynamics} & \cite{datasrc-stanford-network-analysis-project-amazon-may-5-2003-network-dataset} \\ \hline

YouTube friend. net. & $1134890$ & $2987624$ & $93.93$ & $0.18$ & $0.29$ & $0.40$ & $0.36$ & \cite{leskovec2012, konect} & \cite{datasrc-konect-youtube-friend-net-network-dataset} \\ \hline

 \end{tabular}
    \caption{\textbf{Real networks.} Continuation of Table~\ref{tab:real1}}
    \label{tab:real2}
\end{table*}

}

%\bibliography{PRA_PRODUCTION_READY_VERIFIED}{}

%%%%%%%%%%%%%%%%%%%%%%


%%%%%%%%%%%%%%%%%%%%%%

\end{document}